\documentclass[useAMS,usenatbib]{mn2e} 
\usepackage[dvips]{graphicx}
\usepackage[english]{babel}
\usepackage{amsmath}
\usepackage{amssymb}
\usepackage{textcomp}
\usepackage{natbib}
\title[Gas accretion on to galaxies and haloes]{The rates and modes of gas accretion on to galaxies and their gaseous haloes}
\author[F. van de Voort et al.]{Freeke van de Voort$^{1}$\thanks{E-mail: fvdvoort@strw.leidenuniv.nl},
Joop~Schaye$^1$,
C.~M.~Booth$^1$,
Marcel~R.~Haas$^1$, and 
\newauthor
Claudio~Dalla~Vecchia$^{1,2}$\\
$^{1}$Leiden Observatory, Leiden University, Postbus 9513, 2300 RA, Leiden, The Netherlands\\
$^{2}$Max Planck Institute for Extraterrestrial Physics, Giessenbachstra\ss{}e 1, 85748 Garching, Germany
}

\begin{document}

\date{Accepted not yet. Received not yet; in original form \today}

\pagerange{\pageref{firstpage}--\pageref{lastpage}} \pubyear{2010}

\maketitle

\label{firstpage}

\begin{abstract}
We study the rate at which gas accretes on to galaxies and haloes and investigate whether the accreted gas was shocked to high temperatures before reaching a galaxy. For this purpose we use a suite of large cosmological, hydrodynamical simulations from the OWLS project, which uses a modified version of the smoothed particle hydrodynamics code \textsc{gadget-3}. We improve on previous work by considering a wider range of halo masses and redshifts, by distinguishing accretion on to haloes and galaxies, by including important feedback processes, and by comparing simulations with different physics.

Gas accretion is mostly smooth, with mergers only becoming important for groups and clusters. The specific rate of gas accretion on to haloes is, like that for dark matter, only weakly dependent on halo mass. For halo masses $M_\mathrm{halo}\gg10^{11}$~M$_\odot$ it is relatively insensitive to feedback processes. In contrast, accretion rates on to galaxies are determined by radiative cooling and by outflows driven by supernovae and active galactic nuclei. Galactic winds increase the halo mass at which the central galaxies grow the fastest by about two orders of magnitude to $M_\mathrm{halo}\sim10^{12}$~M$_\odot$. 

Gas accretion is bimodal, with maximum past temperatures either of order the virial temperature or $\lesssim10^5$~K. The fraction of gas accreted on to haloes in the hot mode is insensitive to feedback and metal-line cooling. It increases with decreasing redshift, but is mostly determined by halo mass, increasing gradually from less than 10\% for $\sim10^{11}$~M$_\odot$ to greater than 90\% at $\sim10^{13}$~M$_\odot$. In contrast, for accretion on to galaxies the cold mode is always significant and the relative contributions of the two accretion modes are more sensitive to feedback and metal-line cooling. On average, the majority of stars present in any mass halo at any redshift were formed from gas accreted in the cold mode, although the hot mode contributes typically over 10\% for $M_\mathrm{halo}\gtrsim10^{11}$~M$_\odot$.

Thus, while gas accretion on to haloes can be robustly predicted, the rate of accretion on to galaxies is sensitive to uncertain feedback processes. Nevertheless, it is clear that galaxies, but not necessarily their gaseous haloes, are predominantly fed by gas that did not experience an accretion shock when it entered the host halo.

\end{abstract}

\begin{keywords}
galaxies: evolution -- galaxies: formation -- intergalactic medium -- cosmology: theory
\end{keywords}

\section{Introduction}

In the standard cosmological constant or vacuum dominated cold dark matter ($\Lambda$CDM) model mass assembles hierarchically, with the smallest structures forming first. While the collapse of dark matter halts as it reaches virial equilibrium in haloes, baryons can radiate away their binding energy, allowing them to collapse further and fragment into smaller structures, such as stars and galaxies. These galaxies then grow through mergers and gas accretion. 

After virialization, a gas cloud can be in one of three regimes. If the characteristic cooling time exceeds the Hubble time-scale, the gas will not be able to radiate away the thermal energy that supports it and will therefore not collapse. If, on the other hand, the cooling time is smaller than the Hubble time, but larger than the dynamical time-scale, then the cloud can adjust its density and temperature quasi-statically. It will increase both its density and temperature while maintaining hydrostatic equilibrium. Finally, if the cooling time is shorter than the dynamical time, the cloud will cool faster than it can collapse, lowering the Jeans mass and possibly leading to fragmentation. This is the regime in which galaxies are thought to form \citep{Silk1977, Rees1977}. In reality the situation must, however, be more complicated as the density, and thus the cooling and dynamical times, will vary with radius.

Gas falling towards a galaxy gains kinetic energy until it reaches the hydrostatic halo. 
If the infall velocity is supersonic, it will experience a shock and heat to the virial temperature of the halo. According to the simplest picture of spherical collapse, all gas in a dark matter halo is heated to the virial temperature of that halo, reaching a quasi-static equilibrium supported by the pressure of the hot gas. Gas can subsequently cool radiatively and settle into a rotationally supported disc, where it can form stars \citep[e.g.][]{Fall1980}. We call this form of gas accretion \emph{`hot accretion'} \citep{Katz2003, Keres2005}.

Within some radius, the so-called cooling radius, the cooling time of the gas will, however, be shorter than the age of the Universe. If the cooling radius lies well inside the halo, which is the case for high-mass haloes, a quasi-static, hot atmosphere will form. Accretion on to the galaxy is then regulated by the cooling function. If, on the other hand, this radius is larger than the virial radius, then there will be no hot halo and the gas will not go through an accretion shock at the virial radius. Because gas accreted in this manner may never have been heated to the virial temperature, we refer to this mode of accretion as \emph{`cold accretion'} \citep{Katz2003, Keres2005}. The rapid cooling of gas in low-mass haloes was already shown by \citet{Rees1977} and \citet{White1978}. The accretion rate on to the central galaxy then depends on the infall rate, but not on the cooling rate \citep{White1991}. Simulations confirmed the existence of gas inside haloes which was never heated to the virial temperature of the halo \citep{Katz1991, Kay2000, Fardal2001, Katz2003}.

The cooling time $t_\mathrm{cool}$ is shorter at higher redshift, because the density, $\rho$, is higher and $t_\mathrm{cool}\propto\rho^{-1}$. The Hubble time $t_\mathrm{H}$ is also shorter, but with a weaker dependence, $t_\mathrm{H}\propto \rho^{-1/2}$. Hence, the mode of gas accretion on to galaxies will depend on redshift, with cold accretion more prevalent at higher redshifts for a fixed virial temperature.
The mode of gas accretion will also depend on halo mass. Higher-mass haloes have higher virial temperatures and thus, at least for $T\gtrsim10^6$~K, longer cooling times. Hence, hot accretion will dominate for the most massive haloes.
The dominant form of accretion will thus depend on both the mass of the halo and on the accretion redshift \citep{Katz2003, Birnboim2003, Keres2005, Ocvirk2008, Keres2009a, Brooks2009, Crain2010}.

Although both analytic and semi-analytic studies of galaxy formation usually assume spherical symmetry \citep[e.g.][]{Binney1977, White1991, Birnboim2003}, numerical simulations show different geometries. As the matter in the Universe collapses, it forms a network of sheets and filaments, the so-called `cosmic web'. Galaxies form in the densest regions, the most massive galaxies where filaments intersect. These structures can have an important effect on gas accretion. If a galaxy is being fed along filaments, the average density of the accreting gas will be higher. The cooling time will thus be smaller and it will be easier to radiate away the gravitational binding energy. Filaments may therefore feed galaxies preferentially through cold accretion.
Both modes can coexist. Especially at high redshift, cold streams penetrate the hot virialized haloes of massive galaxies \citep{Keres2005, Dekel2006, Ocvirk2008, Keres2009a, Dekel2009a}.

Galaxy colours and morphologies are observed to be roughly bimodal. They can be divided into two main classes: blue, star forming spirals and red, passive ellipticals \citep[e.g.][]{Kauffmann2003, Baldry2004}. The latter tend to be more massive and to reside in clusters. It has been suggested by \citet{Dekel2006} that this observed bimodality is caused by the two different mechanisms for gas accretion. If, for example, feedback from active galactic nuclei (AGN) were to prevent the hot gas from cooling, massive galaxies with little cold accretion would have very low star formation rates. 
If, on the other hand, cold accretion flows were less susceptible to such feedback, then low-mass haloes could host discs of cold gas, which could form stars efficiently. However, this cannot be the whole story, because suppressing star formation from gas accreted in the hot mode is not nearly sufficient to reproduce the observations \citep{Keres2009b}.

In this work we use a large number of cosmological, hydrodynamical simulations
from the OverWhelmingly Large Simulations project
\citep[OWLS;][]{Schaye2010} to investigate accretion rates, the separation of hot and cold modes, and their dependence on halo mass and redshift. Our work extends earlier work in several ways. 
By combining simulations with different box sizes, each of which uses at least as many particles as previous simulations, we are able to cover a large dynamic range in halo masses and a large range of redshifts. 
While earlier work used only a single physical model, our use of a range of models allows us to study the role of metal-line cooling, supernova (SN) feedback, and AGN feedback. Some of these processes had been ignored by earlier studies. For example, \citet{Keres2005, Keres2009a, Keres2009b} ignored metal-line cooling and SN feedback. Using a semi-analytic model, \citet{Benson2010} emphasized the importance of including the effect of feedback. \citet{Brooks2009} did not include metal-line cooling, but did include SN feedback. \citet{Ocvirk2008} included both metal-line cooling and weak SN feedback, but could only study the high-redshift behaviour as their simulation was stopped at $z\approx1.5$. AGN feedback was not included by any previous study, although \citet{Khalatyan2008} did show the existence of cold accretion in a simulation of a single group of galaxies with AGN feedback.
In contrast to previous work, we distinguish between accretion on to haloes and accretion on to galaxies (i.e.\ the interstellar medium).  
We find that while hot mode accretion dominates the growth of high-mass haloes, cold mode accretion is still most important for the growth of the galaxies in these haloes. The different physical models give similar results for accretion on to haloes, implying that the results are insensitive to feedback processes, but the inclusion of such processes is important for accretion on to galaxies.

\begin{table*}
\caption{\label{tab:res} \small Simulation parameters: simulation identifier, comoving box size ($L_\mathrm{box}$), number of dark matter particles ($N$, the number of baryonic particles is equal to the number of dark matter particles), mass of dark matter particles ($m_\mathrm{DM}$), initial mass of gas particles ($m_\mathrm{gas}$), final simulation redshift, number of resolved haloes (containing more than 250 dark matter particles) at $z=2$ ($N_\mathrm{halo}(z=2)$), and number of resolved haloes (containing more than 250 dark matter particles) at $z=0$ ($N_\mathrm{halo}(z=0)$).}
\begin{tabular}[t]{lrlllrrr}
\hline
\hline \\[-3mm]
simulation & $L_\mathrm{box}$ & $N$ & {$m_\mathrm{DM}$} & {$m_\mathrm{gas}$} & $z_\mathrm{final}$ & $N_\mathrm{halo}(z=2)$ & $N_\mathrm{halo}(z=0)$ \\
 & (h$^{-1}$Mpc) & & {(M$_\odot$)} & {(M$_\odot$)} \\
\hline \\[-4mm]
\emph{L100N512} & 100 & 512$^3$ & $5.56\times10^8$ & $1.19\times10^8$ & 0 & 10913 & 18463 \\
\emph{L100N256} & 100 & {256$^3$} & $4.45\times10^9$ & $9.84\times10^9$ & 0 & 838 & 2849 \\
\emph{L050N512} & {50} & 512$^3$ & $6.95\times10^7$ & $1.48\times10^7$ & 0 & 12648 & 15884  \\
\emph{L025N512} & {25} & 512$^3$ & $8.68\times10^6$ & $1.85\times10^6$ & 2 &12768 & - \\
\emph{L025N256} & {25} & {256$^3$} & $6.95\times10^7$ & $1.48\times10^7$ & 2 & 1653 & - \\
\emph{L025N128} & {25} & {128$^3$} & $5.56\times10^8$ & $1.19\times10^8$ & 0 & 159 & 338 \\
\hline
\end{tabular}
\end{table*}

This paper is organized as follows. The simulations are described in Section~\ref{sec:sim}, including all the model variations. In Section~\ref{sec:Trho} we describe some properties of the gas in the temperature-density plane for our fiducial simulation. The ways in which haloes are identified and gas accretion is determined is discussed in Section~\ref{sec:acc}. The accretion rates for all simulations can be found in Section~\ref{sec:accrate}, for accretion on to haloes as well as for accretion on to galaxies. In Section~\ref{sec:halo} we compare hot and cold accretion on to haloes for the different simulations and we do the same for accretion on to galaxies in Section~\ref{sec:sf}. Finally, we compare our results with previous work in Section~\ref{sec:discuss} and summarize our conclusions in Section~\ref{sec:conclude}. 

\section{Simulations} \label{sec:sim}

To investigate the temperature distribution of accreted gas, we use a modified version of \textsc{gadget-3} \citep[last described in][]{Springel2005}, a smoothed particle hydrodynamics (SPH) code that uses the entropy formulation of SPH \citep{Springel2002}, which conserves both energy and
entropy where appropriate. This work is part of the OWLS \citep{Schaye2010} project, which consists of a large number of cosmological simulations, with varying (subgrid) physics. We first summarize the subgrid prescriptions for the reference simulation. The other simulations are described in Section \ref{sec:var}. As the simulations are fully described in \citet{Schaye2010}, we will only summarize their main properties here.

The cosmological simulations described here assume a $\Lambda$CDM cosmology with parameters as determined from the Wilkinson microwave anisotropy probe year 3 (WMAP3) results, $\Omega_\mathrm{m} = 1 - \Omega_\Lambda = 0.238$, $\Omega_\mathrm{b} = 0.0418$, $h = 0.73$, $\sigma_8 = 0.74$, $n = 0.951$. These values are consistent\footnote{The only significant discrepancy is in $\sigma_8$, which is 8\% lower than the value favoured by the WMAP 7-year data.} with the WMAP year~7 data \citep{Komatsu2011}. 

A cubic volume with periodic boundary conditions is defined, within which the mass is distributed over $N^3$ dark matter and as many gas particles. The box size (i.e.\ the length of a side of the simulation volume) of the simulations used in this work is 25, 50, or 100~$h^{-1}$Mpc, with $N=512$, unless stated otherwise. The (initial) particle masses for baryons and dark matter are $1.2\times10^8(\frac{L_\mathrm{box}}{100\ \mathrm{h^{-1}Mpc}})^3(\frac{N}{512})^{-3}$~M$_\odot$ and $5.6\times10^8(\frac{L_\mathrm{box}}{100\ \mathrm{h^{-1}Mpc}})^3(\frac{N}{512})^{-3}$~M$_\odot$, respectively, and are listed in Table~\ref{tab:res}, as are the number of resolved haloes at $z=2$ and $z=0$. We use the notation \emph{L***N\#\#\#}, where \emph{***} indicates the box size and \emph{\#\#\#} the number of particles per dimension.
The gravitational softening length is 7.8~$(\frac{L_\mathrm{box}}{100\ \mathrm{h^{-1}Mpc}})(\frac{N}{512})^{-1}$~$h^{-1}$kpc comoving, i.e.\ 1/25 of the mean dark matter particle separation, but we imposed a maximum of 2~$(\frac{L_\mathrm{box}}{100\ \mathrm{h^{-1}Mpc}})(\frac{N}{512})^{-1}$~$h^{-1}$kpc proper.

The primordial abundances are $X = 0.752$ and $Y = 0.248$, where $X$ and $Y$ are the mass fractions of hydrogen and helium, respectively. The abundances of eleven elements (hydrogen, helium, carbon, nitrogen, oxygen, neon, magnesium, silicon, sulphur, calcium, and iron) released by massive stars (type II SNe and stellar winds) and intermediate mass stars (type Ia SNe and asymptotic giant branch stars) are followed as described in \citet{Wiersma2009b}.
We assume the stellar initial mass function (IMF) of \citet{Chabrier2003}, ranging from 0.1 to 100~M$_\odot$. As described in \citet{Wiersma2009a}, radiative cooling and heating are computed element-by-element in the presence of the cosmic microwave background radiation and the \citet{Haardt2001} model for the UV/X-ray background from galaxies and quasars.

Star formation is modelled according to the recipe of \citet{Schaye2008}.
The Jeans mass cannot be resolved in the cold, interstellar medium (ISM), which could lead to artificial fragmentation \citep[e.g.][]{Bate1997}.
Therefore a polytropic equation of state $P_\mathrm{tot}\propto\rho_\mathrm{gas}^{4/3}$ is implemented for densities exceeding $n_\mathrm{H}=0.1$~cm$^{-3}$, where $P_\mathrm{tot}$ is the total pressure and $\rho_\mathrm{gas}$ the density of the gas. It keeps the Jeans mass fixed with respect to the gas density, as well as the ratio of the Jeans length and the SPH smoothing kernel.
Gas particles with proper densities $n_\mathrm{H}\ge0.1$~cm$^{-3}$ and temperatures $T\le10^5$~K are moved on to this equation of state and can be converted into star particles. 
The star formation rate (SFR) per unit mass depends on the gas pressure and is set to reproduce the observed Kennicutt-Schmidt law \citep{Kennicutt1998}.

\begin{table*}
\caption{\label{tab:owls} \small Simulation parameters: simulation identifier, cooling including metals (Z cool), wind velocity ($v_\mathrm{wind}$), wind mass loading ($\eta$), and AGN feedback included (AGN). Differences from the reference model are indicated in bold face. The last column gives the box size and particle number, from Table \ref{tab:res}, used with these simulation parameters ($L_\mathrm{box}$ and $N$).}
\begin{tabular}[t]{lcrcrr}
\hline
\hline \\[-3mm]
simulation & Z cool & $v_\mathrm{wind}$ & $\eta$  & AGN & $L_\mathrm{box}$ and $N$ \\
  & & (km/s) & & &\\
\hline \\[-4mm]
\emph{REF} & yes & 600 & 2 & no & all listed in Table \ref{tab:res} \\
\emph{NOSN} & yes & \textbf{\ \ \ 0} & \textbf{0}  & no & \emph{L100N512} at $z=2$ \\
\emph{NOSN\_NOZCOOL} & \textbf{no} & \textbf{\ \ \ 0} & \textbf{0} & no & \emph{L100N512, L025N512} \\
\emph{NOZCOOL} & \textbf{no} & 600 & 2 & no & \emph{L100N512, L025N512} \\
\emph{WML4} & yes & 600 & \textbf{4} & no & \emph{L100N512, L025N512} \\
\emph{WML1V848} & yes & \textbf{848} & \textbf{1} & no & \emph{L100N512, L025N512} \\
\emph{DBLIMF-} & yes & \textbf{600 \& 1618} & 2 & no & \emph{L100N512, L025N512} \\
\emph{ \ \ CONTSFV1618} &  &  &  &  \\
\emph{WDENS} & yes &  \multicolumn{2}{r}{\textbf{density dependent}} & no & \emph{L100N512, L025N512} \\
\emph{AGN} & yes & 600 & 2 & \textbf{yes} & \emph{L100N512, L025N512} \\
\hline
\end{tabular}
\end{table*}

Feedback from star formation is implemented using the prescription of \citet{Vecchia2008}. About 40 percent of the energy released by type II SNe is injected locally in kinetic form. 
The rest of the energy is assumed to be lost radiatively. Each gas particle within the SPH smoothing kernel of the newly formed star particle has a probability of being kicked.
For the reference model, the mass loading parameter $\eta = 2$, meaning that on average the total mass of the particles being kicked is twice the mass of the star particle formed. Because the winds sweep up surrounding material, the effective mass loading can be higher. The initial wind velocity is 600~km/s for the reference model, which is consistent with observations of starburst galaxies, both locally \citep{Veilleux2005} and at high redshift \citep{Shapley2003}. \citet{Schaye2010} showed that these parameter values yield a peak, global star formation rate density that agrees with observations.

The simulation data is saved at discrete output redshifts with interval $\Delta z=0.125$ at $0\le z\le 0.5$, $\Delta z=0.25$ at $0.5< z\le 4$, and $\Delta z=0.5$ at $4< z\le 9$. This is the time resolution used for determining accretion rates.

\subsection{Model variations} \label{sec:var}

To see whether or not our results are sensitive to specific physical processes or subgrid prescriptions, we have performed a suite of simulations in which many of the simulation parameters are varied. These are listed in Table~\ref{tab:owls}.

The importance of metal-line cooling can be demonstrated by comparing the reference simulation (\emph{REF}) to a simulation in which primordial abundances are assumed when calculating the cooling rates (\emph{NOZCOOL}).
Similarly, the effect of including SN feedback can be studied by comparing a simulation without SN feedback (\emph{NOSN}) to the reference model. Because the metals cannot be expelled without feedback, they pile up, causing efficient cooling and star formation. To limit the cooling rates, we performed a simulation in which both cooling by metals and feedback from SNe were ignored (\emph{NOSN\_NOZCOOL}). Simulation \emph{NOSN\_L025N512} was stopped just below $z=4$ and the $z=0.125$ snapshot is missing for \emph{NOSN\_L100N512}. We therefore cannot always show results for this simulation.

In massive haloes the pressure of the ISM is too high for winds with velocities of 600~km/s to blow the gas out of the galaxy \citep{Vecchia2008}. Keeping the wind energy per unit stellar mass constant, we increased the wind velocity by $\sqrt{2}$ to $v_\mathrm{wind}=848$~km/s, while halving the mass loading to $\eta=1$ (\emph{WML1V848}). To enable the winds to eject gas from haloes with even higher masses, the velocity and mass loading can be scaled with the local sound speed, while keeping the energy injected per unit stellar mass constant (\emph{WDENS}). 

To investigate the dependence on SN energy given to the ISM, we ran a simulation using almost all of the energy available from SNe. The wind mass loading $\eta=4$, a factor of two higher than in the reference simulation. The wind velocity is the same, $v_\mathrm{wind}=600$~km/s (\emph{WML4}).

There is some evidence that the IMF is top-heavy in extreme environments, like starburst galaxies \citep[e.g.][]{Padoan1997, Klessen2007, Dabringhausen2009}.
We have performed a simulation in which the IMF is top-heavy for stars formed at high pressures, $P_\mathrm{tot}/k_\mathrm{B}>1.7\times10^5$~cm$^{-3}$K, where $k_\mathrm{B}$ is Boltzmann's constant, and therefore in high-mass haloes. The critical pressure was chosen such that $\sim$10\% of the stars are formed with a top-heavy IMF. It results in more SNe per solar mass of stars formed and therefore stronger winds. More energy is therefore put into the winds blown by these star particles, $v_\mathrm{wind}=1618$~km/s and $\eta=2$ (\emph{DBLIMFCONTSFV1618}).

Finally, we have included AGN feedback (\emph{AGN}). Black holes inject 1.5\% of the rest-mass energy of the accreted gas into the surrounding matter in the form of heat. The model is described and tested in \citet{Booth2009}, who also demonstrate that it reproduces the observed mass density in black holes and the observed scaling relations between black hole mass and both central stellar velocity dispersion and and stellar mass. \citet{McCarthy2010} have shown that our model AGN reproduces the observed stellar mass fractions, SFRs, stellar age distributions and thermodynamic properties of galaxy groups.

\begin{figure*}
\center
\includegraphics[scale=0.5]{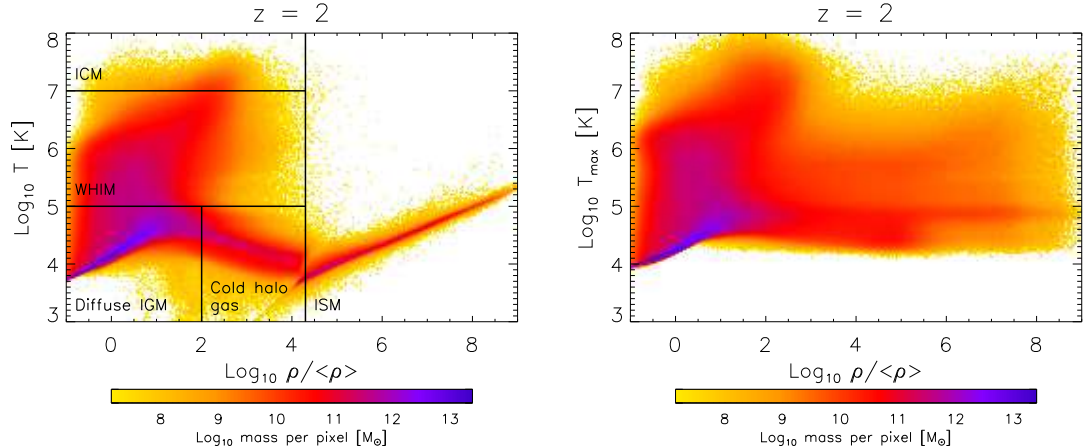}
\caption {\label{fig:gas_T_maxT_rho} Current temperature (left) and maximum past temperature (right) against current overdensity for gas particles at $z=2$ in the reference simulation in a 50~$h^{-1}$Mpc box. The logarithm of the total gas mass in a pixel is used for colour coding. The black lines separate the different phases of the gas. Gas with $T\gtrsim10^5$~K has been shock heated. The $T-\rho$ relation for cold, low-density gas ($T\lesssim10^5$~K and $\rho\lesssim10\langle\rho\rangle$ is set by heating by the UV background and adiabatic cooling. At $\rho/\langle\rho\rangle>10^{1.5}$ radiative cooling dominates over cooling by the expansion of the Universe. Gas with $n_\mathrm{H}>0.1$~cm$^{-3}$ ($\rho/\langle\rho\rangle>10^{4.3}$ at $z=2$) is assumed to be part of the unresolved, multiphase ISM and is put on an effective equation of state. The `temperature' of this gas merely reflects the pressure of the multiphase ISM and is therefore not used to update $T_\mathrm{max}$. The scatter in the equation of state is caused by adiabatic cooling of inactive particles in between time steps. Gas that is condensing on to haloes or inside haloes has $T_\mathrm{max}\gtrsim10^{4.5}$~K, which reflects the peak in the T-$\rho$ relation (at $\rho/\langle\rho\rangle>10^{1.5}$, $T\sim10^{4.5}$~K) visible in the left panel.}
\end{figure*}

\subsection{Maximum past temperature}

The Lagrangian nature of the simulation is exploited by tracing each fluid element back in time, which is particularly convenient for this project, in which we are studying the temperature history of accreted gas. During the simulations the maximum temperature, $T_\mathrm{max}$, and the redshift at which it was reached, $z_\mathrm{max}$, were stored in separate variables. The variables were updated for each SPH particle at every time step for which the temperature was higher than the previous maximum temperature. The artificial temperature the particles obtain when they are on the equation of state (i.e.\ when they are part of the unresolved multiphase ISM) was ignored in this process. This may cause us to underestimate the maximum past temperature of gas that experienced an accretion shock at higher densities. Ignoring such shocks is, however, consistent with our
aims, as we are interested in the maximum temperature reached
\emph{before} the gas entered the galaxy.

Shock heating could be missed if the particle heats and cools rapidly in a single time step. 
In SPH simulations, a shock is smeared out over a few smoothing lengths, leading to in-shock cooling \citep{Hutchings2000}. We find the average value for the infall velocity to be $\sim 5\times 10^2~\bigl( \frac{M}{10^{13}~M_\odot}\bigr)^{1/3} (1+z)^{1/2}$~km/s. On average the SPH smoothing length at halo accretion is $\sim 10^2$~kpc (comoving) for all \emph{L100N512} simulations and at all redshifts. The particle will take $10^8~\bigl( \frac{M}{10^{13}~M_\odot}\bigr)^{-1/3} (1+z)^{-3/2}$~years to traverse this distance. A shock will therefore take at least a few times $10^8$~years. 
In reality the accretion shock will proceed almost instantaneously, minimizing radiative losses. In the simulations described here gas cools at each time step, so also while it is being shocked. The finite spatial resolution can thus result in lower maximum temperatures than the actual post-shock temperatures. If this effect were important, increasing the resolution would increase the hot fraction, i.e.\ the fraction of the gas that is accreted in the hot mode, because the accretion shocks would be less broadened. The opposite is the case, as we will show in Figure~\ref{fig:ratio2p0}.

Even with infinite time resolution, the post-shock temperatures may, however, not be well defined. Electrons and protons will temporarily have different temperatures in the post-shock gas, because they differ in mass, an effect that we have not included. It will take a short while before they equilibrate through collisions or plasma effects. Another effect, which is also not included in our simulations, is that shocks may be preceded by radiation from the shock, which may affect the temperature evolution.

\section{The temperature-density distribution} \label{sec:Trho}

The left-hand panel of Figure~\ref{fig:gas_T_maxT_rho} shows the mass-weighted distribution of gas in the temperature-density plane at $z=2$ for simulation \emph{REF\_L050N512}. Gas with densities up to $\sim10^2$ times the cosmic average density represents the diffuse intergalactic medium (IGM). A significant fraction of this gas resides in filamentary structures. It can be heated to temperatures above $10^5$~K when kinetic energy, generated by gravitational infall or galactic winds, is converted into thermal energy. We refer to this tenuous, shock-heated gas as the warm-hot intergalactic medium (WHIM). The intracluster medium (ICM) is the very hot $T\gtrsim10^7$~K gas located in galaxy groups and clusters. Gas at overdensities $\rho/\langle\rho\rangle\gtrsim10^2$, but much lower temperatures ($T\sim10^4$~K) resides mostly in filaments and low-mass haloes.

Most of the gas is located in the purple region, with $\rho/\langle\rho\rangle<10$.
The temperature of this gas is determined by the combination of photoheating by the UV background and adiabatic cooling by the expansion of the Universe. Although the slope of this temperature-density relation is close to adiabatic, it is actually determined by the temperature dependence of the recombination rate \citep{Hui1997}. 

The turnover density, above which the typical gas temperatures decrease, occurs when radiative cooling starts to dominate over adiabatic cooling. The overdensity at which this happens depends on redshift. At $z=2$ it occurs at $\rho/\langle\rho\rangle \approx 10^{1.5}$.
The distribution of the WHIM (broad red region, with $T>10^5$~K) is set in part by the cooling function, to which especially heavier atoms, like oxygen, contribute. In particular, the lack of dense gas ($\rho/\langle\rho\rangle\gtrsim10^3$) with $10^5$~K~$\lesssim T\lesssim10^7$~K is due to radiative cooling.

Gas with proper hydrogen number density $n_\mathrm{H}>0.1$~cm$^{-3}$, corresponding at this redshift to overdensity $\rho/\langle\rho\rangle>10^{4.3}$, represents the ISM.
This high-density gas is put on an equation of state if its temperature was below $10^5$~K when it crossed the density threshold, because the cold and warm phases of this dense medium are not resolved by the simulations. Therefore, the temperature merely reflects the imposed effective pressure and the density should be interpreted as the mean density of the unresolved multi-phase ISM.
The spread in the temperature-density relation on the equation of state is caused by the adiabatic extrapolation of inactive particles between time steps.
In addition, the relation is broadened by differences in the mean molecular weight, $\mu$, of the gas, which depends on the density, temperature and elemental abundances of the gas. 

The right-hand panel of Figure~\ref{fig:gas_T_maxT_rho} shows the maximum past temperature reached at $z\ge2$, as a function of the $z=2$ baryonic overdensity.
All dense gas has reached temperatures of $\gtrsim10^{4.5}$~K at some point. Dense ($\rho/\langle\rho\rangle > 10^{1.5}$) gas cannot have $T_\mathrm{max}$ much below $10^{4.5}$~K, because of photoheating\footnote{Exceptions are gas that reached high densities before reionization, which happens at $z=9$ in our simulations, and gas with very high metallicities.}. 
The maximum past temperature tends to remain constant once the gas has reached densities $\gtrsim10^2\langle\rho\rangle$, resulting in the horizontal trend in the figure. There is no dense $\rho/\langle\rho\rangle\sim10^4$gas at the highest maximum temperatures (above 10$^7$~K), because the cooling time of gas that is heated to this temperature at lower densities ($\rho/\langle\rho\rangle\sim10^{2-3}$) is longer than a Hubble time, preventing it from cooling and reaching higher densities.

The maximum past temperature reached for dense gas depends on whether the gas has been heated to the virial temperature of its halo and on whether it has been shock-heated by galactic winds. 

\section{Defining gas accretion} \label{sec:acc}

To see how haloes accrete gas, we first need to find and select the haloes. This can be done in several different ways and we will discuss three of them. Although we choose to use the one based on the gravitational binding energy, our results are insensitive to the halo definition we use. Finally, we link haloes in two subsequent snapshots in order to determine which gas has entered the haloes.

\subsection{Identifying haloes and galaxies}
The first step towards finding gravitationally bound structures is to identify dark matter haloes. These can for example be found using a Friends-of-Friends (FoF) algorithm. If the separation between two dark matter particles is less than 20\% of the average separation (the linking length $b=0.2$), they are placed in the same group. Because the particles all have the same mass, a fixed value of $b$ will correspond to a fixed overdensity at the boundary of the group of $\rho/\langle\rho\rangle\approx60$ \citep[e.g.][]{Frenk1988}. Assuming a radial density profile $\rho(r)\propto r^{-2}$, corresponding to a flat rotation curve, such a group has an average overdensity of $\langle\rho_\mathrm{halo}\rangle/\langle\rho\rangle\approx 180$ \citep[e.g.][]{Lacey1994}, close to the value for a virialized object predicted by the top-hat spherical collapse model \citep[e.g.][]{Padmanabhan2002}. The minimum number of dark matter particles in a FoF group is set to 25. Many of the smallest groups will not be gravitationally bound. Baryonic particles are placed in a group if their nearest dark matter neighbour is part of the group.

Problems arise because unbound particles can be attached to a group, physically distinct groups can be linked by a small (random) particle bridge, and because substructure within a FoF halo is not identified. We use \textsc{subfind} \citep{Springel2001, Dolag2009}
on the FoF output to find the gravitationally bound particles and to identify subhaloes. The properties of gas that is being accreted are expected to depend on the properties of the parent (or main) halo in which the subhaloes are embedded. We therefore only look at accretion on to these main haloes and exclude gas accretion by satellites.

Another way of defining haloes is to use a spherical overdensity criterion. The radius $R_\mathrm{vir}$, centred on the most bound particle of a FoF halo, is found within which the average density agrees with the prediction of the top-hat spherical collapse model in a $\Lambda$CDM cosmology \citep{Bryan1998}.
All the particles within $R_\mathrm{vir}$ are then included in the halo.

We chose to use only main haloes identified by \textsc{subfind}, but we have checked that our results do not change significantly when using FoF groups or spherical overdensities instead.

Except for Figures~\ref{fig:acchaloeos}, \ref{fig:ratio2p0}, and \ref{fig:haloeosstar}, we include only main haloes containing more than 250 dark matter particles in our analysis of halo accretion. This corresponds to a minimum total halo mass of $M_\mathrm{halo}\approx10^{11.2}$~M$_\odot$ in the 100~$h^{-1}$Mpc box, $10^{10.3}$~M$_\odot$ in the 50~$h^{-1}$Mpc box,  and $10^{9.4}$~M$_\odot$ in the 25~$h^{-1}$Mpc box.
For these limits our mass functions agree very well with the \citet{Sheth1999} fit.

For each resolved halo, we identify the ISM of the central galaxy with the star forming (i.e.\ $n_\mathrm{H}>0.1$~cm$^{-3}$) gas particles in the main halo which are inside 15\% of the virial radius \citep{Sales2010}. We use 15\% of the virial radius to exclude small star forming substructures in the outer halo, which are not identified by \textsc{subfind}. Gas accretion on to satellite galaxies is excluded.

Because the galaxy is much smaller than its parent halo, it is not as well resolved. When investigating accretion on to galaxies (as opposed to haloes), we therefore impose a 1000 dark matter particle limit, corresponding to a minimum total halo mass of $M_\mathrm{halo}\approx10^{11.8}$~M$_\odot$ in the 100~$h^{-1}$Mpc box, $10^{10.9}$~M$_\odot$ in the 50~$h^{-1}$Mpc box, and $10^{10.0}$~M$_\odot$ in the 25~$h^{-1}$Mpc box. The exceptions are Figure~\ref{fig:acchaloeos} and \ref{fig:haloeosstar}, which show accretion rates and hot fractions down to 1 dex below this limit. See Section \ref{sec:accrate} for more discussion on convergence.

\subsection{Selecting gas particles accreted on to haloes}

We select gas particles accreted on to haloes as follows.
For each halo at $z=z_2$ (which we will also refer to as `the descendant') we identify its progenitor at the previous output redshift $z_1>z_2$. We determine which halo contains most of the descendant's 25 most bound dark matter particles and refer to this halo as `the progenitor'. 
If the fraction of the descendant's 25 most bound particles that was not in any halo at $z_1$ is greater than the fraction that was part of the progenitor, then we discard the halo from our analysis, which rarely happens above our resolution threshold. If two or more haloes contain the same number of those 25 particles, we select the one that contains the dark matter particle that is most bound to the descendant.

We identify those particles that are in the descendant, but not in its progenitor as having been accreted at $z_2\le z<z_1$. The accreted particles have to be gaseous at $z_1$, i.e.\ before they were accreted, but can be either gaseous or stellar at redshift $z_2$. The accreted gas can have densities exceeding the star formation threshold, in which case it cannot obtain a higher maximum past temperature. Gas can be accreted multiple times.

To distinguish mergers from smooth accretion, we exclude accreted particles that reside in (sub)haloes above some maximum mass at redshift $z_1$. We would like to have a criterion that is not directly dependent on resolution, so that the same objects are included in runs with different particle numbers if the simulations are converged with respect to resolution. We therefore set this maximum allowed halo mass to 10\% of the descendant's (main halo) mass. Thus, smooth accretion excludes mergers with a mass ratio greater than 1:10. We experimented with different thresholds and the results are insensitive to this choice.

\subsection{Selecting gas particles accreted on to galaxies}

We consider particles that are part of the ISM or stellar at $z_2$, and that were gaseous but not part of the ISM at $z_1$, to have been accreted on to a galaxy at $z_2\le z<z_1$. Gas can be accreted multiple times. Accretion on to the ISM and accretion on to a galaxy are the same in this study and these terms are used interchangeably. 

In this way, accretion through galaxy mergers is automatically excluded, because the gas that is part of the ISM of another galaxy at $z_1$ is excluded. We can allow for mergers by identifying those particles that are in the galaxy at $z_2$, but were not in the progenitor galaxy at $z_1$. 

\section{Total gas accretion rates} \label{sec:accrate}

For brevity, accretion between, for example, $z=2.25$ and $z=2$ will be referred to as accretion just before $z=2$.

\begin{figure}
\center
\includegraphics[scale=0.5]{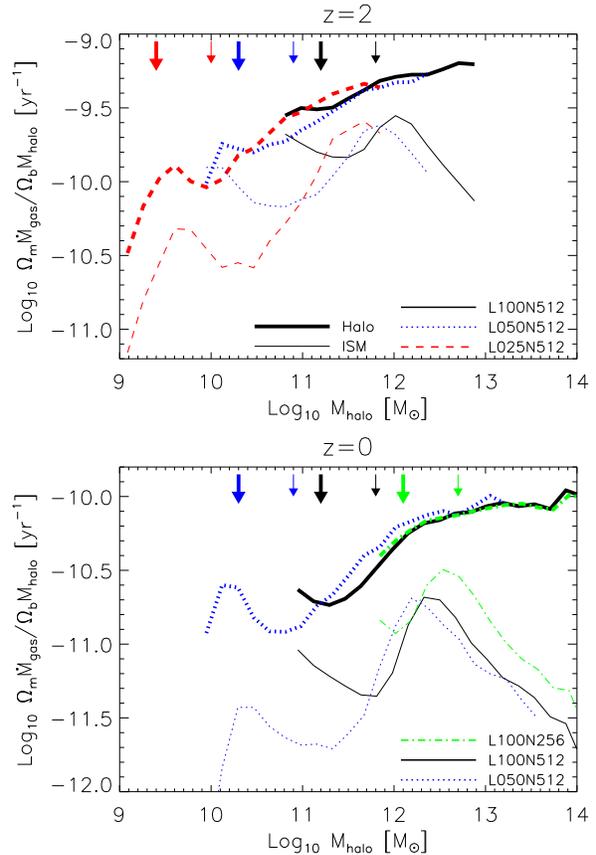}
\caption {\label{fig:acchaloeos} Specific gas smooth accretion rates on to \textit{haloes} (higher, thick curves) and central \textit{galaxies} (lower, thin curves) against total halo mass at $z=2$ (top panel) and $z=0$ (bottom panel). The normalization for halo accretion gives an estimate of the time it would take a halo to grow to its current mass at the current accretion rate. The same is not true for galaxy accretion, since we use the same normalization as for halo accretion. Both panels show simulations spanning a factor~64 in mass resolution. Each mass bin contains at least 10~haloes. Big (small) arrows correspond to the adopted resolution limit for accretion on to haloes (galaxies) for \emph{L025N512} (red), \emph{L050N512} (blue), \emph{L100N512} (black) and \emph{L100N256} (green). The specific halo accretion rate is converged and increases slightly with halo mass. The specific galaxy accretion rate is not fully converged at $z=2$, but the convergence is better at $z=0$. The galaxy accretion rate increases with halo mass for $M_\mathrm{halo}<10^{12}$~M$_\odot$ and decreases with halo mass for higher halo masses. It is much smaller than the halo accretion rate.}
\end{figure}

The average total gas accretion rate of a halo depends on its mass. Figure~\ref{fig:acchaloeos} shows the average specific accretion rates $\Omega_\mathrm{m}\dot{M}_\mathrm{gas}/\Omega_\mathrm{b}M_\mathrm{halo}$, where $\Omega_\mathrm{m}$ and $\Omega_\mathrm{b}$ are the matter and baryon density parameters, respectively, as a function of halo mass for $z=2$ (top panel) and $z=0$ (bottom panel) for haloes containing at least 100 dark matter particles for various simulations using the reference model. Thick curves show the specific accretion rates on to haloes. We have divided the gas accretion rate by the total halo mass and baryon fraction $\Omega_\mathrm{b}/\Omega_\mathrm{m}$, so that the normalization gives an estimate of the time it would take a halo to grow to its current mass with the current accretion rate. 

The thin curves in Figure~\ref{fig:acchaloeos} show specific smooth accretion rates on to galaxies against halo mass.
While the inverse of the specific halo accretion rate equals the time it takes to grow the halo at its current accretion rate, the same is not true for the specific galaxy accretion rate, because we divide by the halo mass rather than by the galaxy mass. This definition allows us, however, to directly compare the two accretion rates.

Comparing the three thick curves in each panel, we see that the specific halo accretion rate is converged with the numerical resolution. It increases with halo mass, especially at low halo masses. The specific galaxy accretion rate (thin curves) is not fully converged at $z=2$, but the convergence improves at $z=0$ (compare the two highest resolutions, dotted and solid curve, at the high mass end). Below  $M_\mathrm{halo}<10^{12}$~M$_\odot$ the specific galaxy accretion rate increases more steeply with halo mass than the specific halo accretion rate. Above this halo mass, the specific galaxy accretion rate decreases steeply, whereas the rate keeps increasing for haloes. The accretion rate on to galaxies is always much lower than the halo accretion rate and this difference is larger at $z=0$. Hence, only a small fraction of the gas accreted on to haloes ends up in galaxies and the efficiency of galaxy formation is highest in haloes with $M_\mathrm{halo}\approx10^{12}$~M$_\odot$.

The dynamic range covered by these simulations is very large, because we use different box sizes. The resolution increases with decreasing box size. We have performed box size convergence tests (at fixed resolution), but do not show them as the convergence with box size is excellent for all halo masses. Increasing the resolution shows that the halo accretion rates are not fully converged around the 100~particle resolution threshold adopted in Figure~\ref{fig:acchaloeos}. In this regime, halo mergers cannot always be identified. In the rest of this paper, we therefore set the minimum halo mass for accretion on to haloes to correspond to 250~dark matter particles, as indicated by the big arrows. 

The galaxy accretion rates diverge at the low-mass end. This is expected, because haloes with 100 dark matter particles will have very few star forming gas particles and because 15\% of the virial radius is close to the spatial resolution limit. We will therefore only include the haloes where 15\% of the virial radius is larger than 5 times the softening length and therefore consisting of at least 1000 dark matter particles, as indicated by the small arrows.
The galaxy accretion rates at the high-mass end are not completely converged at $z=2$. The accretion rate decreases by up to a factor of two if the mass resolution is increased by a factor of 64. With increasing resolution, more gas reaches higher densities. If this gas becomes star forming before accreting on to the central galaxy, it is not included in the smooth accretion rate. If we instead include galaxy mergers (not shown), which increases the accretion rate for high-mass galaxies slightly, then the results are in fact fully converged for the 50 and 25~$h^{-1}$Mpc box. 

The bottom panel shows that the convergence is better at $z=0$.

Although the results of these convergence tests are encouraging, we caution the reader that we cannot exclude the possibility that higher resolution simulations would show larger differences. For example, \citet{Vecchia2008} found that higher resolution is required to obtained converged predictions for the galactic winds. On the other hand, it is not clear the resolution requirements inferred by \citet{Vecchia2008} carry over to the present simulations, because they used idealised, isolated disk galaxies haloes that started with thinner stellar disks than are formed in our simulations and their gas disks were initially embedded in a vacuum.

\subsection{Accretion on to haloes}

\begin{figure}
\center
\includegraphics[scale=0.5]{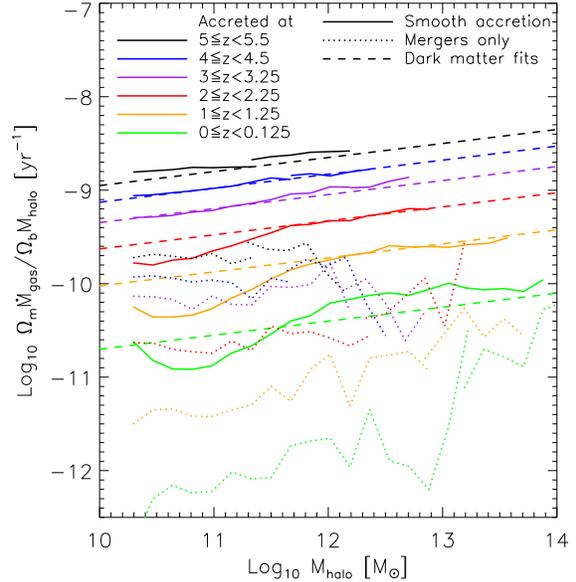}
\caption {\label{fig:accratez} Specific gas accretion rates on to haloes for redshifts $z=5$ (black, top curve) to $z=0$ (green, bottom curve) against the total halo mass at the redshift of accretion for the simulation \emph{REF\_L050N512}. We added the highest mass bins from \emph{REF\_L100N512} to extend the dynamic range. The solid curves are for smooth accretion, whereas the dotted curves show the specific accretion rate due to mergers with mass ratios greater than 1:10. Each mass bin contains at least 10 haloes. Dashed lines show fits to dark matter accretion rates \citep{Dekel2009b, Neistein2006}. The specific smooth accretion rate decreases with redshift and increases only mildly with halo mass. Except for clusters ($M_\mathrm{halo}\ge10^{14}$M$_\odot$) at $z=0$, halo growth is dominated by smooth accretion.}
\end{figure}

Figure~\ref{fig:accratez} shows the average specific gas accretion rate on to haloes as a function of halo mass for $z=5$ (black, top curve) to $z=0$ (green, bottom curve) for run \emph{L050N512}. We also show the high-mass bins from \emph{L100N512}, which are not sampled by \emph{L050N512}, because of its smaller volume.
The solid curves show the rates for smooth accretion, the dotted curves for mergers with mass ratios greater than 1:10. The specific smooth accretion rate decreases with decreasing redshift and increases slowly with halo mass. The accretion rate, as opposed to the \textit{specific} accretion rate, is thus roughly proportional to the halo mass.

Our accretion rates are generally in quantitative agreement with other simulations \citep{Ocvirk2008, Keres2005}. They also agree well with a fit to an analytic prediction for dark matter accretion rates based on the extended Press-Schechter formalism given by \citet{Dekel2009b} and derived by \citet{Neistein2006}. This fit is shown by the dashed lines. For low-mass haloes ($M_\mathrm{halo}\le10^{12}$~M$_\odot$) at $0\le z\le2$ the accretion rate is lower than predicted. This suppression is due to SN winds, as we will discuss below. Without SN feedback, the gas accretion rates follow the dark matter accretion rates also for low-mass haloes at low redshift.

The specific gas accretion rate through mergers is much lower than the specific smooth accretion rate, except for high-mass haloes ($M_\mathrm{halo}\gtrsim10^{14}$~M$_\odot$) at $z=0$. The dominance of smooth accretion is consistent with recent work on \emph{dark matter} accretion, which has shown that mergers with mass ratios greater than 1:10 contribute less than 20-30\% to the total halo accretion and at least 10-40\% of the accretion is genuinely diffuse \citep{Fakhouri2010, Angulo2010, Genel2010, Wang2010}.

\begin{figure*}
\center
\includegraphics[scale=0.58]{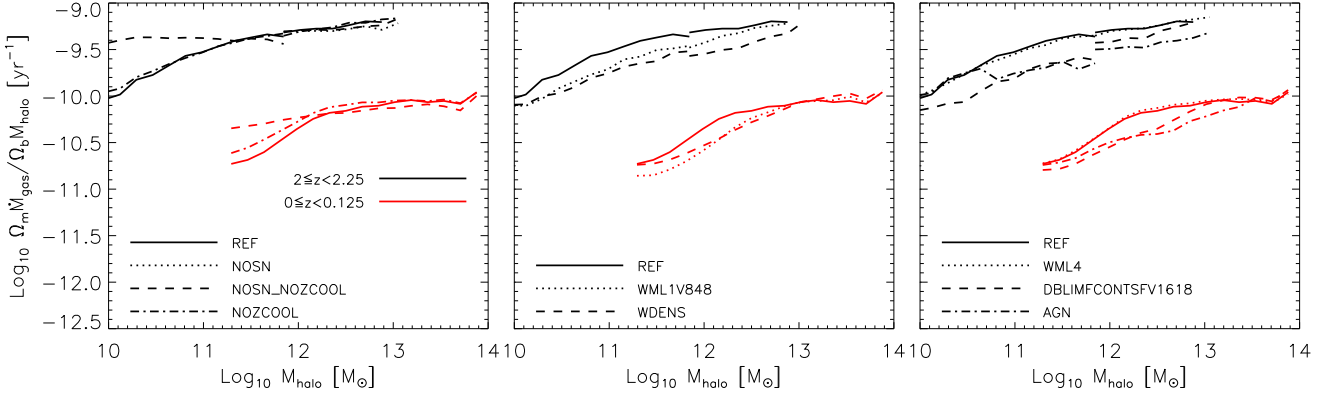}
\caption {\label{fig:diffaccrate} Specific gas smooth accretion rates on to \textit{haloes} against total halo mass at $z=2$ (top, black curves) and at $z=0$ (bottom, red curves) for different simulations. The normalization gives an estimate of the time it would take a halo to grow to its current mass at the current accretion rate. The curves at low (high) halo masses are for simulations in a  25 (100)~$h^{-1}$Mpc box using 2$\times 512^3$~particles. The solid curves use a simulation with the reference parameters (\emph{REF}) and are repeated in all panels. The different simulations are described in Section~\ref{sec:var}. Each mass bin contains at least 10 haloes. Each halo contains at least 1000 dark matter particles for the $z=2$ curves. This is higher than our resolution limit for accretion on to haloes, but it removes the overlap between simulations of different resolution. Each halo at $z=0$ contains at least 250 dark matter particles. \textit{Left panel:} Turning off feedback from SNe results in up to 0.6~dex higher accretion rates, for low-mass haloes. \textit{Middle panel:} Increasing the wind velocity causes the specific accretion rate to decrease by up to 0.2~dex, over the mass range where the feedback is efficient. \textit{Right panel:} Efficient feedback from a top-heavy IMF or from AGN even reduces the accretion rates for the highest halo masses, although the differences between these models are still small, at most 0.3 dex.}
\end{figure*}
\begin{figure*}
\center
\includegraphics[scale=0.58]{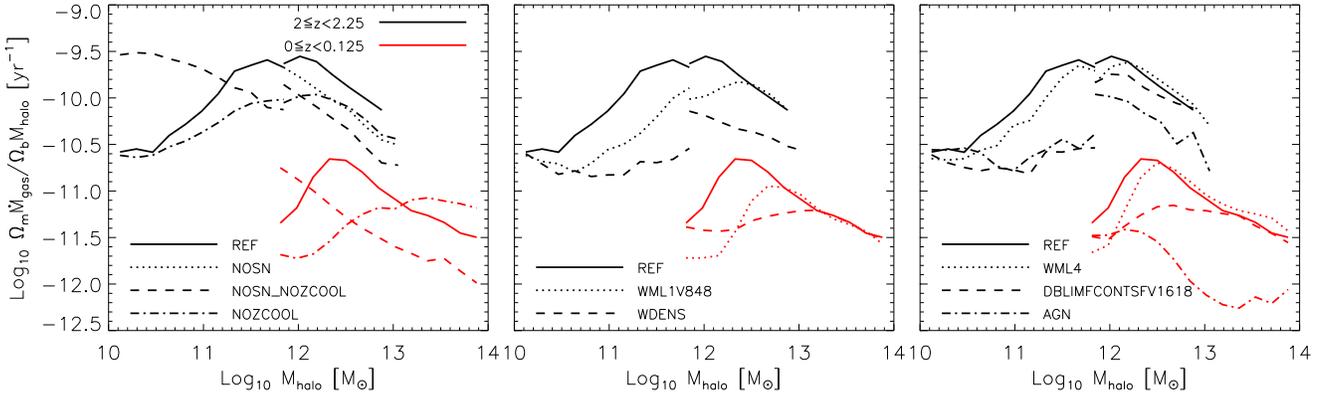}
\caption {\label{fig:diffaccrateeos} Specific gas smooth accretion rates on to \textit{galaxies} against halo mass for different simulations. The curves at low (high) halo masses are derived from simulations in a  25 (100)~$h^{-1}$Mpc box using $2\times512^3$ particles. The line styles, colours, and normalization are identical to those used in Figure~\ref{fig:diffaccrate}. The normalization does \emph{not} give an estimate for the time it would take the galaxy to grow, because we divide by the halo mass, not the galaxy mass. Each mass bin contains at least 10 haloes. In the absence of SN feedback the specific accretion rate on to galaxies declines with halo mass, indicating that gas accretes less efficiently on to galaxies in higher-mass haloes. Leaving out metal-line cooling decreases the accretion rate most strongly for $M_\mathrm{halo}\sim 10^{12}$~M$_\odot$. Efficient SN feedback reduces the accretion rates substantially for galaxies in low-mass haloes, resulting in a peak in the specific gas accretion rate at $M_\mathrm{halo}\sim10^{12}$~M$_\odot$. The effects of feedback and metal-line cooling are much stronger for accretion on to galaxies than for accretion on to haloes and can result in differences of an order of magnitude.}
\end{figure*}

The power of the OWLS project lies in the fact that many simulations have been performed with different prescriptions for physical processes such as cooling and feedback. They have been described in Section~\ref{sec:sim}. Figure~\ref{fig:diffaccrate} shows how the different physical processes affect the specific halo accretion rates at $z=2$ (black curves) and at $z=0$ (red curves). At $z=2$, we include results from both the \emph{L025N512} and \emph{L100N512} versions of each model in order to extend the range of halo masses. These simulations have different box sizes and differ by a factor of 64 in mass resolution. For some models the run with lower resolution and larger box size, which is used for the high halo masses, is not fully converged, resulting in discontinuities. Only the \emph{L100N512} simulations were run down to $z=0$. 

The first thing to notice is the fact that the halo smooth accretion rates are very similar for all models. For the simulations with strong winds (\emph{DBLIMFCONTSFV1618}) or AGN feedback (\emph{AGN}) the rates are smaller than those for the reference model by up to 0.3-0.4~dex. The other models differ even less, except for the run without SN feedback (\emph{NOSN\_NOZCOOL}) which predicts 0.6~dex higher accretion rates at the lowest halo masses. The differences are similar at the two redshifts.

The left panel of Figure~\ref{fig:diffaccrate} shows the effect of excluding metal-line cooling and SN feedback. The simulations without SN feedback (\emph{NOSN}, which is only available for \emph{L100N512} at $z=2$, and \emph{NOSN\_NOZCOOL}) have nearly completely flat specific accretion curves. As is the case for dark matter accretion rates, there is only a small increase with halo mass (see Figure~\ref{fig:accratez}). Including SN feedback with $v_\mathrm{wind}=600$~km/s (\emph{REF} and \emph{NOZCOOL}) reduces the accretion rates for low-mass haloes by up to 0.6~dex (at $M_\mathrm{halo}=10^{10}$~M$_\odot$). It is possible that the true accretion rates are the same, but some gas is pushed out of the halo before a snapshot is made and is therefore not counted as having accreted. A more likely possibility is, however, that galactic winds prevented the gas around low-mass haloes from accreting. Hence, more gas remains available for accretion on to massive galaxies which explains why those have somewhat higher accretion rates if SN feedback is included. Excluding metal-line cooling has no effect on the halo accretion rates. This indicates that halo accretion rates are not set by cooling.

SN feedback models with higher wind velocities, but using the same amount of energy per unit stellar mass, plotted in the middle panel of Figure~\ref{fig:diffaccrate}, give lower accretion rates over the range of masses for which the winds are able to eject gas. Higher wind velocities keep the feedback efficient up to higher halo masses \citep[][Haas et al. in preparation]{Vecchia2008, Schaye2010}.

The right panel shows the results for models that use more energy for feedback than the reference model. A top-heavy IMF in starbursts (\emph{DBLIMFCONTSFV1618}) reduces the accretion rates by up to 0.3~dex.
AGN feedback (\emph{AGN}) reduces the accretion rates by up to 0.4~dex for $M_\mathrm{halo}\gtrsim10^{11}$~M$_\odot$. 

In summary, metal-line cooling does not significantly change the halo accretion rates, but very efficient feedback from stars and/or AGN can suppress the accretion rates by factors of a few.

\subsection{Accretion on to galaxies}

The three lower, thin curves in the top and bottom panels of Figure~\ref{fig:acchaloeos} show specific smooth accretion rates on to galaxies (i.e.\ the ISM) against total halo mass at $z=2$ and 0, respectively. The specific accretion rate peaks at $M_\mathrm{halo}\approx10^{12}$~M$_\odot$. At $z=0$ the peak occurs at slightly higher halo masses. For both redshifts the peak falls at $T_\mathrm{vir}\sim10^6$~K, close to the bump in the cooling curve due to iron \citep{Wiersma2009a}. At higher halo masses, and thus higher virial temperatures, cooling times become long, preventing shocked gas from condensing on to galaxies. At lower halo masses, feedback prevents gas from entering the ISM. For $M_\mathrm{halo}\approx10^{12}$~M$_\odot$ the galaxy accretion rate is about a factor of two lower than the halo accretion rate, but the difference is much larger for other halo masses. 

Figure~\ref{fig:diffaccrateeos} shows the same as Figure~\ref{fig:diffaccrate}, but for accretion on to galaxies. 
For models \emph{WDENS}, \emph{DBLIMFCONTSFV1618}, and \emph{AGN} the convergence is poor for the low-resolution runs, as can be seen from the discontinuities. Because the feedback in these models depends on the gas density, they are sensitive to the resolution \citep{Schaye2010}. Note, however, that the difference in mass resolution is enormous, a factor of 64, and that the high-resolution model may thus be much closer to convergence than the comparison suggests. Nevertheless, we caution the reader that the accretion rates may be different for higher resolution simulations. 

The variations in the feedback prescriptions result in similar differences as for accretion on to haloes, although they are generally much larger. Excluding metal-line cooling gives slightly different results than halo accretion.

\begin{figure}
\center
\includegraphics[scale=0.5]{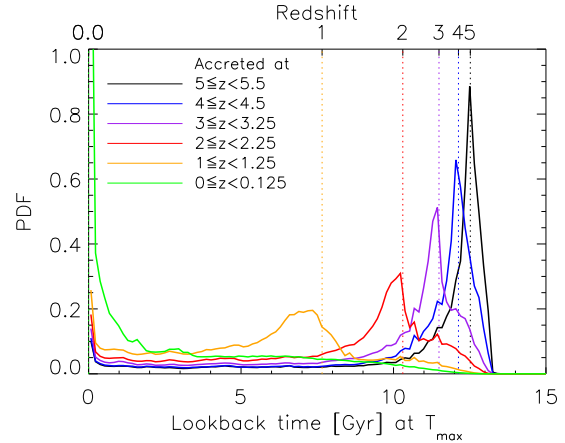}
\caption {\label{fig:lbt} PDF of the lookback time at which the maximum past temperature (evaluated at $z=0$) was reached for gas accreted just before $z=5$, 4, 3, 2, 1, and 0 in black, blue, purple, red, orange, and green, respectively, combining \emph{REF\_L050N512} and \emph{REF\_L100N512}. The top $x$-axis indicates the corresponding redshift. The maximum temperature reached by a gas particle is associated with the accretion event, as evidenced by the fact that the PDFs peak at the accretion redshift.}
\end{figure}

For $M_\mathrm{halo}\lesssim10^{13}$~M$_\odot$ less gas reaches the star formation threshold (i.e.\ accretes on to galaxies) without metal-line cooling (\emph{NOZCOOL}), particularly for $M_\mathrm{halo}\sim10^{12}$~M$_\odot$ at $z=2$ and $M_\mathrm{halo}\sim10^{12.4}$~M$_\odot$ at $z=2$. The reduction is less than 0.4~dex at $z=2$, but a full order of magnitude at $z=0$. For $M_\mathrm{halo}\gtrsim10^{13}$~M$_\odot$ more gas accretes on to galaxies at $z=0$ if metal-line cooling is excluded. Because there is less cooling, less gas accretes on to low-mass galaxies and at high redshift. Therefore, there is more gas left to accrete on to high mass galaxies at low redshift.

For simulations without SN feedback the specific galaxy accretion rate peaks at $M_\mathrm{halo}\sim10^{10}$~M$_\odot$, which is two orders of magnitude lower than when SN feedback is included. Without SN feedback, the galaxy accretion rates are a bit \emph{lower} for high halo masses. This could be because much of the gas that accreted on to a halo at higher redshift has in that case already been accreted on to the galaxy's progenitors. Alternatively, the galaxy accretion rates could be higher in the presence of SN feedback due to the increased importance of recycling. If winds are able to blow gas out of the galaxy, but not out of the halo, as may be the case for high-mass haloes, then the same gas elements may be accreted on to the galaxy more than once \citep{Oppenheimer2010}. For $M_\mathrm{halo}\lesssim10^{11}$~M$_\odot$ at $z=2$ and $M_\mathrm{halo}\lesssim10^{12}$~M$_\odot$ at $z=0$, on the other hand, the accretion rates are higher, because more gas accretes on to the halo and because there is no feedback to stop halo gas from accreting on to the galaxy.

Galaxy accretion rates are thus much more sensitive to metal-line cooling and to SN and AGN feedback than halo accretion rates. The difference between models can be as large as 1~dex. Feedback processes determine the halo mass for which galaxy formation is most efficient.

\section{Hot and cold accretion on to haloes} \label{sec:halo}

\begin{figure*}
\center
\includegraphics[scale=.8]{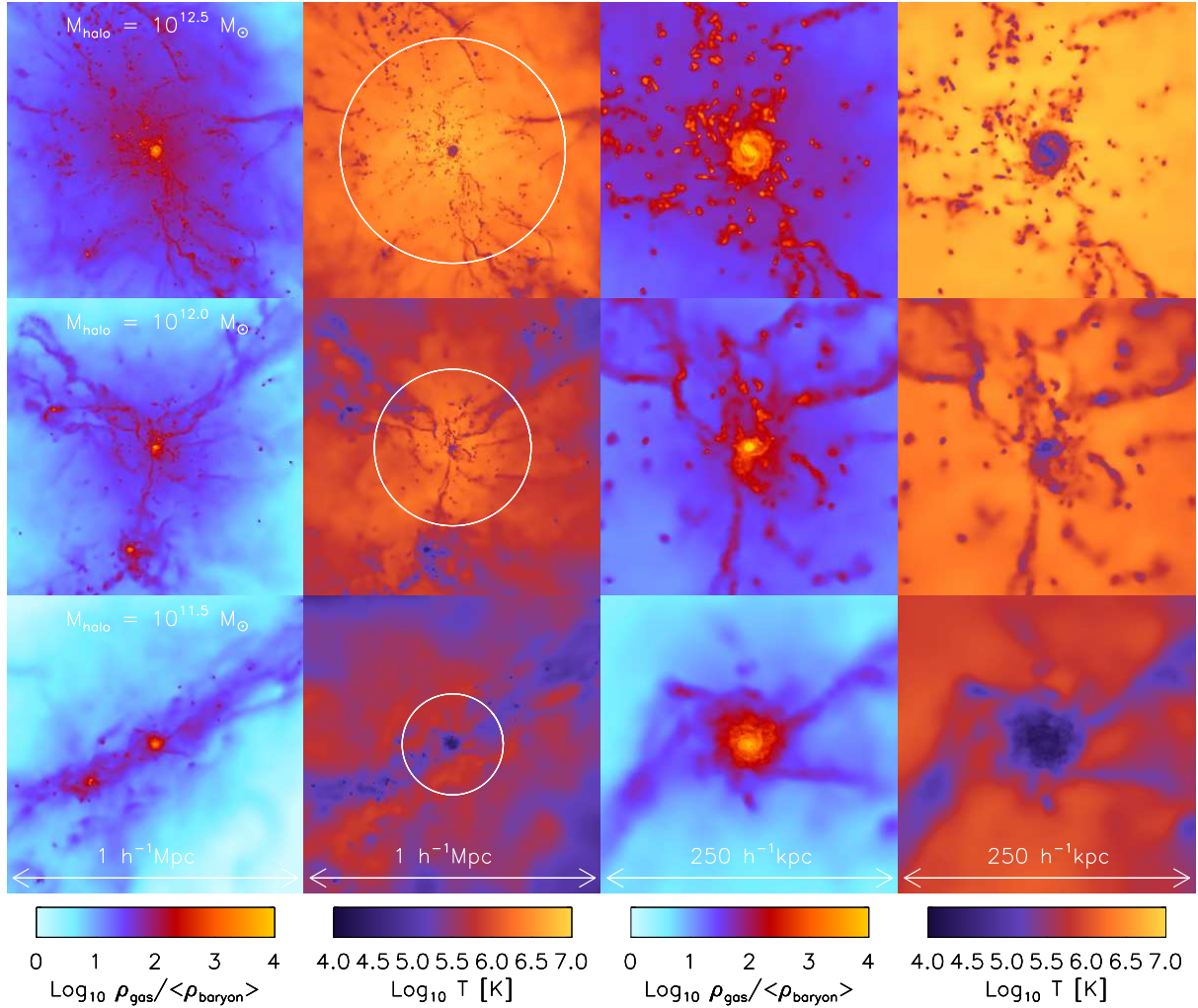}
\caption {\label{fig:single} Gas overdensity (first and third columns) and temperature (second and fourth columns) in a cubic region of 1~$h^{-1}$ comoving Mpc (first and second columns) and 250~$h^{-1}$ comoving kpc (third and fourth columns) centred on haloes of $M_\mathrm{halo}\approx10^{12.5}$, $10^{12}$, and $10^{11.5}$~M$_\odot$ (from top to bottom) at $z=2$ for simulation \emph{REF\_L025N512}. The white circles indicate the virial radii of the haloes, as computed using the overdensity criterion from \citet{Bryan1998}. Cold, dense streams bring gas to the centre. The temperature of the hot gas increases with halo mass. Hot accretion dominates for high-mass haloes, cold accretion for low-mass haloes. The galaxies in the centres of these haloes are discs, surrounded by cold gas. This cold gas is in clumps ($M_\mathrm{halo}\approx10^{12.5}$), disrupted streams ($M_\mathrm{halo}\approx10^{12}$), or smooth streams ($M_\mathrm{halo}\approx10^{11.5}$).}
\end{figure*}

Figure~\ref{fig:lbt} demonstrates that most of the gas reaches its maximum temperature around the time it is accreted on to a halo. Here the probability density function (PDF) of the lookback time at which the gas reaches its maximum temperature, evaluated at $z=0$, is shown for baryonic particles that were accreted as gas particles on to haloes at different redshifts. The vertical dotted lines show the times at which the gas was accreted on to a halo. The fact that the PDFs peak around the accretion redshifts shows that the maximum temperature is usually related to the accretion event.

Some of the gas reaches its maximum temperature significantly later than the redshift at which it was accreted. When two galaxies merge, gas can shock to higher temperatures. Gas can also be affected by winds resulting from SN feedback. Because we are primarily interested in gas accretion, we will from now on evaluate the maximum past temperature of the gas at the first available output redshift after accretion.

Gas can be accreted cold on to haloes with well developed virial shocks if its density is sufficiently high, as can for example be the case in filaments. To illustrate this we show in the left two columns of Figure~\ref{fig:single} the gas overdensity and the temperature in a cubic region of 1~$h^{-1}$~Mpc (comoving) centred on three example haloes at $z=2$ taken from the full sample of 12768 haloes in the high-resolution reference simulation (\emph{REF\_L025N512}). The haloes have total masses $M_\mathrm{halo}\approx10^{12.5}$, $10^{12}$, and $10^{11.5}$~M$_\odot$ (from top to bottom), corresponding to comoving virial radii of 379, 264, and 170~$h^{-1}$kpc, respectively, shown as white circles in the temperature plots. Their virial temperatures are $T_\mathrm{vir}\approx10^{6.3}$, $10^{6.0}$, and $10^{5.7}$~K, respectively. The colour scales are the same for all three haloes. We can immediately see that the average temperature increases with halo mass. Hot gas, heated either by accretion shocks or SN feedback, extends to several virial radii. Without SN feedback, the hot gas would trace the virial radius more accurately for the two lowest mass haloes, as can be seen for the $10^{12}$~M$_\odot$ halo in Figure~\ref{fig:diffsimzoom}. 

Most of the gas around the $10^{12.5}$~M$_\odot$ halo has been heated to temperatures above $10^6$~K and the halo will get most of its gas through hot accretion. Cold streams do penetrate the virial radius, but they seem to break up as they get close to the centre. Even so, a number of small, dense clumps do survive and remain relatively cold. Such cold clumps originating from filamentary gas were also studied by \citet{SommerLarsen2006} and \citet{KeresHernquist2009}.

The $10^{12}$~M$_\odot$ halo is located at the intersection of three filaments. The gas in the filaments is denser and colder than in the surrounding medium. The cold streams become narrower as they get closer to the centre. They are compressed by the high pressure, shock-heated gas around them \citep{Keres2009a}. Cold streams bring gas directly and efficiently to the inner halo.

The $10^{11.5}$~M$_\odot$ halo is embedded in a single filament. It has the lowest virial temperature, so the hot gas is much colder than in the highest mass halo. Cold streams are most prominent and broadest in this halo. This halo will get most of its gas through cold accretion.

The right two columns of Figure~\ref{fig:single} shows zooms of the central 250~$h^{-1}$ kpc (comoving). The $10^{12.5}$~M$_\odot$ halo contains a large number of cold clumps. The galaxies in the $10^{12}$ and $10^{11.5}$~M$_\odot$ haloes are being fed by cold, dense streams. All of these galaxies have formed discs. The galaxy in the $10^{12.5}$~M$_\odot$ halo has clear spiral arms and a bar-like structure. The galaxy in the $10^{12}$~M$_\odot$ halo has a very small disc with cold gas around it, which looks more disturbed. The galaxy in the $10^{11.5}$~M$_\odot$ halo is fairly large with a lot of cold material accreting on to it.

The maximum temperature reached by shock-heated gas is expected to scale with the virial temperature of the halo. However, we do not expect the ratio of $T_\mathrm{max}$ and $T_\mathrm{vir}$ to be exactly unity, because of departures from spherical symmetry, adiabatic compression after virialization, and the factor of a few difference between different definitions of $T_\mathrm{vir}$. 

Dividing $T_\mathrm{max}$ by $T_\mathrm{vir}$ would take out the redshift and halo mass dependence of the virial temperature. We calculate the virial temperature as follows
\begin{align}  \label{eqn:virialtemperature}
T_\mathrm{vir} & = \left(\dfrac{G^2H_0^2\Omega_\mathrm{m}18\pi^2}{54}\right)^{1/3}\dfrac{\mu m_\mathrm{H}}{k_B}M_\mathrm{halo}^{2/3}(1+z), \nonumber \\
           & \approx 3.0\times10^5\ \mathrm{K}\ \left(\dfrac{\mu}{0.59}\right)\left(\dfrac{M_\mathrm{halo}}{10^{12}\ \mathrm{M}_\odot}\right)^{2/3}\left(1+z\right),
\end{align}
where $G$ is the gravitational constant, $H_0$ the Hubble constant, $\mu$ the mean molecular weight, $m_\mathrm{H}$ the mass of a hydrogen atom, and $k_\mathrm{B}$ Boltzmann's constant\footnote{This definition is a factor $2/3$ lower than the virial temperature used by some other authors \citep[e.g.][]{Barkana2001}.}.

\subsection{Dependence on halo mass}

\begin{figure}
\center
\includegraphics[scale=0.5]{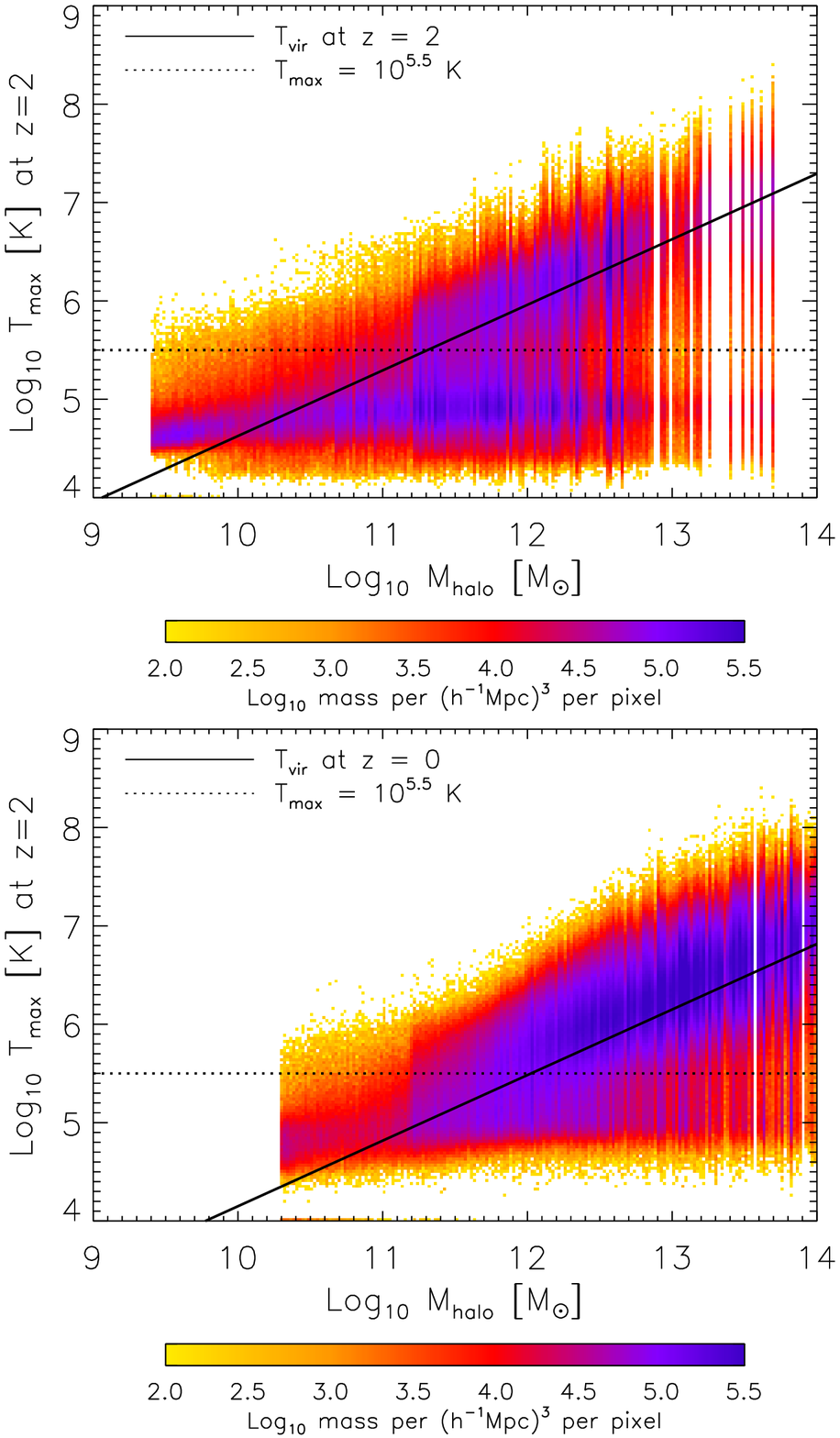}
\caption {\label{fig:maxTmassz2p0} Maximum past temperature at $z=2$ (top panel) and $z=0$ (bottom panel) against total halo mass of the gas smoothly accreted on to haloes for models \emph{REF\_L025N512} and \emph{REF\_L100N512} (top panel) and \emph{REF\_L050N512} and \emph{REF\_L100N512} (bottom panel). The logarithm of the total gas mass per ($h^{-1}$Mpc)$^3$ in a pixel is used for colour coding. The solid line indicates the virial temperature of the halo. The dotted line shows $T_\mathrm{max}=10^{5.5}$~K, where there is a minimum in the mass per pixel at $z=2$. The temperature of gas that is accreted hot scales with the virial temperature. For low-mass haloes, the temperatures of hot and cold accreted gas are comparable. Hot mode accretion is more important for higher halo masses and lower redshifts.}
\end{figure}

Figure~\ref{fig:maxTmassz2p0} illustrates the dependence of the maximum past temperature on halo mass. Shown are scatter plots of the maximum past temperature reached by gas accreting on to haloes just before $z=2$ (top panel) and $z=0$ (bottom panel) against the mass of the halo at these redshifts. The logarithm of the accreted gas mass per ($h^{-1}$Mpc)$^3$ in a pixel is used for colour coding. The virial temperature is indicated by the black line. 

A clear bimodality is visible for accretion at $z=2$, with a minimum at $T_\mathrm{max}\approx10^{5.5}$~K, indicated by the dotted line. This minimum coincides with a maximum at $T\approx10^{5-5.5}$~K in the cooling function \citep[e.g.][]{Wiersma2009a}. 
For gas accreted in the hot mode, which includes most of the gas accreting on to high-mass haloes, the maximum past temperature is within a factor $\approx3$ of the virial temperature and displays the same dependence on mass.
The temperature of the gas accreted in the cold mode is independent of halo mass.

At $z=0$, the lowest $T_\mathrm{max}$ values are higher than at $z=2$. This shift occurs because the density, and hence the cooling rate, increases with redshift. At fixed halo mass, the highest $T_\mathrm{max}$ values are lower at $z=0$, because the virial temperature of a halo at fixed mass decreases with decreasing redshift, as can be seen from Equation \ref{eqn:virialtemperature}. 
For haloes with virial temperatures $T_\mathrm{vir}\lesssim10^5$~K it becomes impossible to tell from this plot whether or not the gas has gone through a virial shock because the virial temperature is similar to the maximum past temperature reached by gas accreting in the cold mode. 
This makes it difficult to separate hot and cold accretion for low-mass haloes at low redshift.
As we will show below, separating hot and cold accretion using a fixed value of $T_\mathrm{max}/T_\mathrm{vir}$ is more difficult than using a fixed value of $T_\mathrm{max}$, because the minimum in the distribution is less pronounced and because it evolves \citep{Keres2005}. In most of this paper, we will therefore use a fixed maximum temperature threshold of $T_\mathrm{max}=10^{5.5}$~K.

\begin{figure}
\center
\includegraphics[scale=0.5]{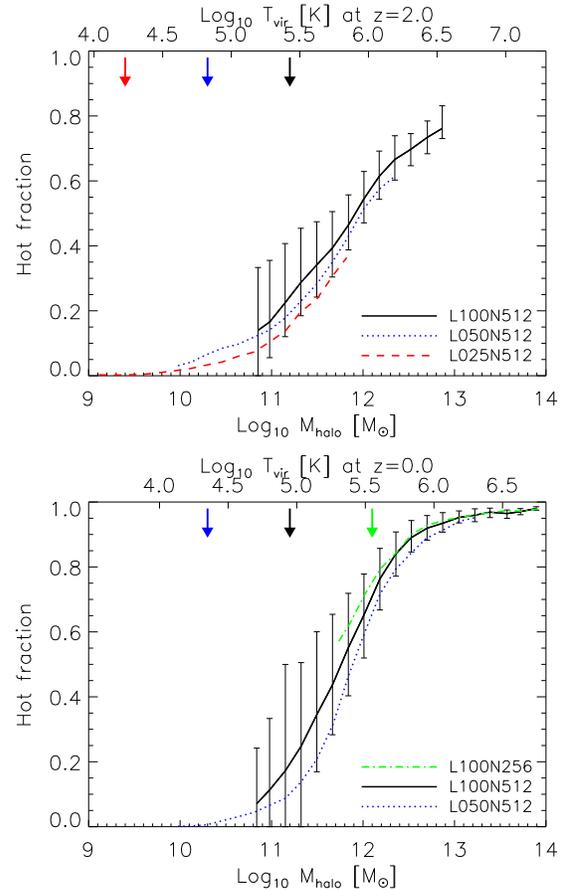}
\caption {\label{fig:ratio2p0} Average fraction of the gas, smoothly accreted on to haloes between $z=2.25$ and $z=2$ (top panel) or between $z=0.125$ and $z=0$ (bottom panel), that has maximum past temperature $T_\mathrm{max}\ge10^{5.5}$~K. The different curves are from simulations of the same reference model but spanning a factor of 64 in mass resolution for each panel. The error bars show the $1\sigma$~ halo to halo scatter for simulation \emph{REF\_L100N512}. Each mass bin contains at least 10 haloes. Arrows correspond to the adopted resolution limit for accretion on to haloes. Cold mode accretion dominates for $M_\mathrm{halo}<10^{12}$~M$_\odot$, but the transition is very gradual.}
\end{figure}
The relative importance of hot accretion increases with halo mass \citep[e.g.][]{Ocvirk2008}. The top panel of Figure~\ref{fig:ratio2p0} shows the fraction of gas smoothly accreting on to haloes in the hot mode, just before $z=2$ for simulations \emph{REF\_L100N512}, \emph{REF\_L050N512}, and \emph{REF\_L025N512}. The bottom panel shows this for accretion just before $z=0$ for simulations \emph{REF\_L100N256}, \emph{REF\_L100N512}, and \emph{REF\_L050N512}.
A particle accreted just before $z=2$ is considered to have been accreted hot if $T_\mathrm{max}(z=2)\ge10^{5.5}$~K. The error bars show the $1\sigma$~halo to halo scatter. 

We have checked, but do not show, that the results are fully converged with box size for fixed resolution. In each panel the three simulations span a factor 64 in mass resolution. The hot fraction decreases slightly with increasing resolution, but the differences are very small. This slight decrease could arise because higher density regions inside clumps and filaments are better sampled with increasing resolution, leading to higher cooling rates in cold gas.

Hot mode accretion becomes indeed more important for higher mass systems. This is expected because only sufficiently massive haloes are capable of providing pressure support for a stable virial shock \citep[e.g.][]{Birnboim2003}. The median hot fraction (not shown) behaves similarly to the mean, although it is smaller for low-mass haloes.

There is no sharp transition from cold to hot accretion. At $z=2$, the hot mode accretion increases from 20\% to 80\% when halo mass increases from 10$^{11}$ to 10$^{13}$~M$_\odot$. The properties of galaxies are therefore not expected to change suddenly at a particular halo mass as has been assumed in some semi-analytic models \citep[e.g.][]{Cattaneo2008, Croton2008}.

\begin{figure}
\center
\includegraphics[scale=0.5]{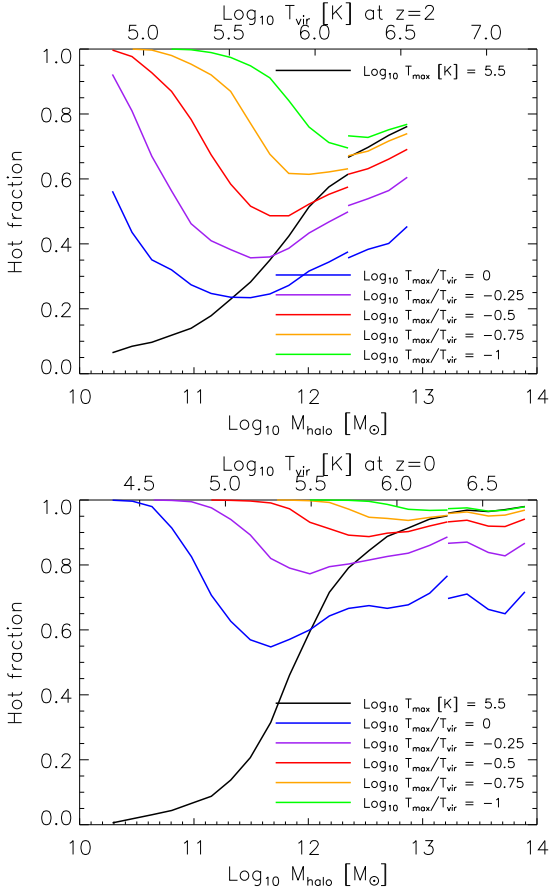}
\caption {\label{fig:virdiff} The average fraction of the gas, accreted on to haloes just before $z=2$ (top panel) or $z=0$ (bottom panel), that has maximum past temperature above a certain temperature threshold. The results are for simulation \emph{REF\_L050N512} with the highest mass haloes from \emph{REF\_L100N512}. The thresholds for the blue, purple, red, orange, and green curves (bottom to top) are $T_\mathrm{max}/T_\mathrm{vir}=1$, $10^{-0.25}$, $10^{-0.5}$, $10^{-0.75}$, and $0.1$, respectively. The black curve shows the average fraction with $T_\mathrm{max}\ge10^{5.5}$~K and is identical to the black curve in Figure~\ref{fig:ratio2p0}. Each mass bin contains at least 10 haloes. The hot fraction depends strongly on the choice we make for the threshold, particularly for lower-mass haloes.}
\end{figure}

In Figure~\ref{fig:virdiff} we compare this result to that obtained when we define hot mode accretion using a maximum temperature threshold that depends on the virial temperature. 
Using $T_\mathrm{max}=T_\mathrm{vir}$ shows that, even for the most massive haloes, only about 40\% of the gas reaches a temperature higher than $T_\mathrm{vir}$ at $z=2$. At $z=0$ about 70\% of the gas accreting on to massive haloes reaches $T_\mathrm{vir}$. Because gas going through a virial shock may only heat to a factor of a few below $T_\mathrm{vir}$, this definition does not discriminate well between hot and cold accretion. If we decrease the critical $T_\mathrm{max}/T_\mathrm{vir}$, then the hot fraction gets close to our previously determined value for the high-mass haloes. For low-mass haloes, however, this results in a sharp upturn of $f_\mathrm{hot}$ as it must approach unity if the threshold maximum temperature falls much below 10$^5$~K.  

The hot fraction therefore depends very much on the definition of the temperature threshold. Depending on the definition, gas accreted cold may in fact have experienced a virial shock.
For haloes with $T_\mathrm{vir}\gg10^5$~K, however, we can safely separate hot and cold accretion and trust the result that a larger fraction of the gas goes through a virial shock for higher-mass haloes.

In the rest of this paper, we will define hot mode accretion using a fixed maximum past temperature threshold of $T_\mathrm{max}=10^{5.5}$~K.

\subsection{Smooth accretion versus mergers}\label{sec:clumpy}

So far we have only looked at `smooth' accretion, i.e.\ we excluded mergers with a mass ratio greater than 1:10. We could have chosen not to exclude mergers, because the gas reservoir of a halo can also grow through mergers. Even though mergers with ratios greater than 1:10 contribute only $\sim10$\% of the total gas accretion for $M_\mathrm{halo}<10^{14}$~M$_\odot$ (see Figure~\ref{fig:accratez}), it is interesting to investigate the differences for hot and cold accretion.

Figure~\ref{fig:z5to0} compares the hot fraction of smooth accretion (solid curves) and all accretion (dotted curves), which takes into account both smooth accretion and mergers. 
The differences are negligibly small. Indeed, the hot fraction for gas accreted in mergers (not shown) is nearly the same as that for gas accreted smoothly.

\subsection{Dependence on redshift}

\begin{figure}
\center
\includegraphics[scale=0.5]{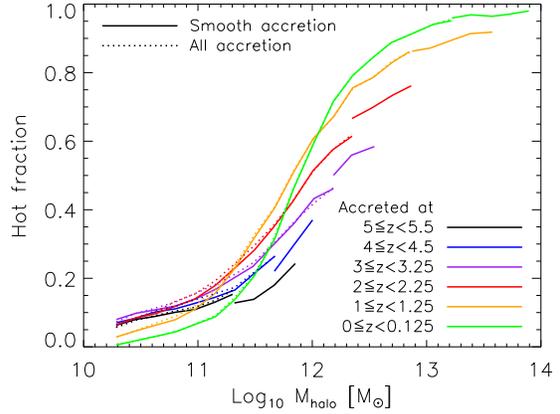}
\caption {\label{fig:z5to0} Hot fraction for accretion on to haloes just before $z=5$, 4, 3, 2, 1, and 0 against halo mass at the same redshift. The curves at low halo masses are obtained from simulation \emph{REF\_L050N512}. At the high-mass end we have added curves for simulation \emph{REF\_L100N512} to extend the dynamic range. Gas is considered to have been accreted hot if it has $T_\mathrm{max}\ge10^{5.5}$. The dotted curves show the hot fraction including both smooth accretion and mergers. Each mass bin contains at least 10 haloes. For a fixed halo mass, hot accretion tends to be more important at lower redshift.}
\end{figure}

The fraction of gas that is accreted on to haloes in the hot mode also depends on redshift, as can be seen from Figure~\ref{fig:z5to0} where the hot fraction is plotted against halo mass for different redshifts. For a given halo mass, the hot fraction increases with time between $z=5$ and $z=1$. Below $z=1$ the rate of evolution slows down, presumably because structure formation slows down, and the sign of the evolution may reverse for low-mass haloes due to the decrease of $T_\mathrm{vir}$ with time. 

At high redshift the proper density of the Universe is higher and the cooling time is therefore shorter ($t_\mathrm{cool}\propto\rho^{-1}\propto(1+z)^{-3}$). The Hubble time is also shorter, but its dependence on redshift is weaker ($t_\mathrm{H}\propto H^{-1}\propto(\Omega_m(1+z)^3+\Omega_\Lambda)^{-1/2}$, so $t_\mathrm{H}\propto(1+z)^{-3/2}$ at high redshift and $t_\mathrm{H}$ is independent of redshift at low redshift). The dynamical time at fixed overdensity has the same redshift dependence as $t_\mathrm{H}$. Hence, cooling is more efficient at higher redshifts \citep[e.g.][]{Birnboim2003} and the hot fraction will thus be lower for a fixed virial temperature. For a fixed halo mass, the evolution in the hot fraction is smaller, because $T_\mathrm{vir}$ is higher at higher redshifts.
The environment can also play a role. At high redshift ($z>2$), $10^{12}$~M$_\odot$ haloes are rare and they tend to reside in highly overdense regions at the intersections of the filaments that make up the cosmic web. Haloes of the same mass are, however, more common at low redshift and form in single filaments, with more average densities. Hence, cold streams may be able to feed massive haloes at high redshift, whereas this may not be possible at low redshift \citep{Keres2005, Dekel2006}.

\begin{figure}
\center
\includegraphics[scale=0.5]{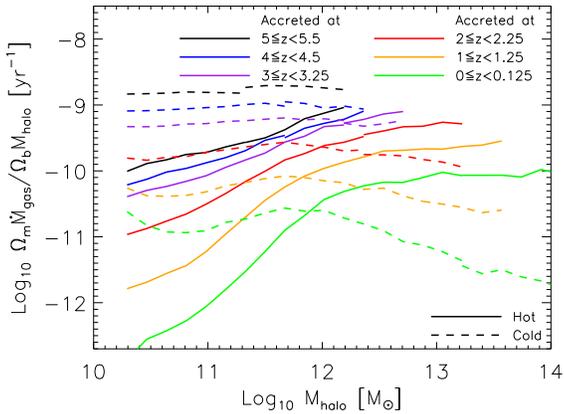}
\caption {\label{fig:specificaccratez} Specific smooth accretion rates of gas on to haloes against total halo mass just before redshifts $z=5$ (black, top curve) to $z=0$ (green, bottom curve) for the simulation \emph{REF\_L050N512}, with the highest mass bins from \emph{REF\_L100N512}. The solid and dashed curves are the rates for hot ($T_\mathrm{max}\ge10^{5.5}$~K) and cold ($T_\mathrm{max}<10^{5.5}$~K) accretion, respectively. Each mass bin contains at least 10 haloes. The specific hot accretion rate increases with halo mass, whereas the specific cold accretion rate decreases for $M_\mathrm{halo}>10^{12}$~M$_\odot$.}
\end{figure}

The specific accretion rate of gas on to haloes increases mildly with halo mass for $0<z<5$, see Figure~\ref{fig:accratez}. Splitting this into hot and cold specific accretion rates reveals a steep increase with halo mass for hot accretion, as shown in Figure~\ref{fig:specificaccratez}. The specific cold accretion rate decreases with halo mass for $M_\mathrm{halo}\gtrsim10^{12}$~M$_\odot$.

From observations we know that the specific star formation rate, i.e.\ the SFR divided by the stellar mass, declines with both time and stellar mass \citep[e.g.][]{Brinchmann2004, Bauer2005, Feulner2005, Chen2008}.  The decline in the specific SFR may be related to the decline in the specific cold accretion rate. However, the decline is much stronger in the observations and present for the entire stellar mass range probed.

\subsection{Effects of physical processes}

\begin{figure}
\center
\includegraphics[scale=0.685]{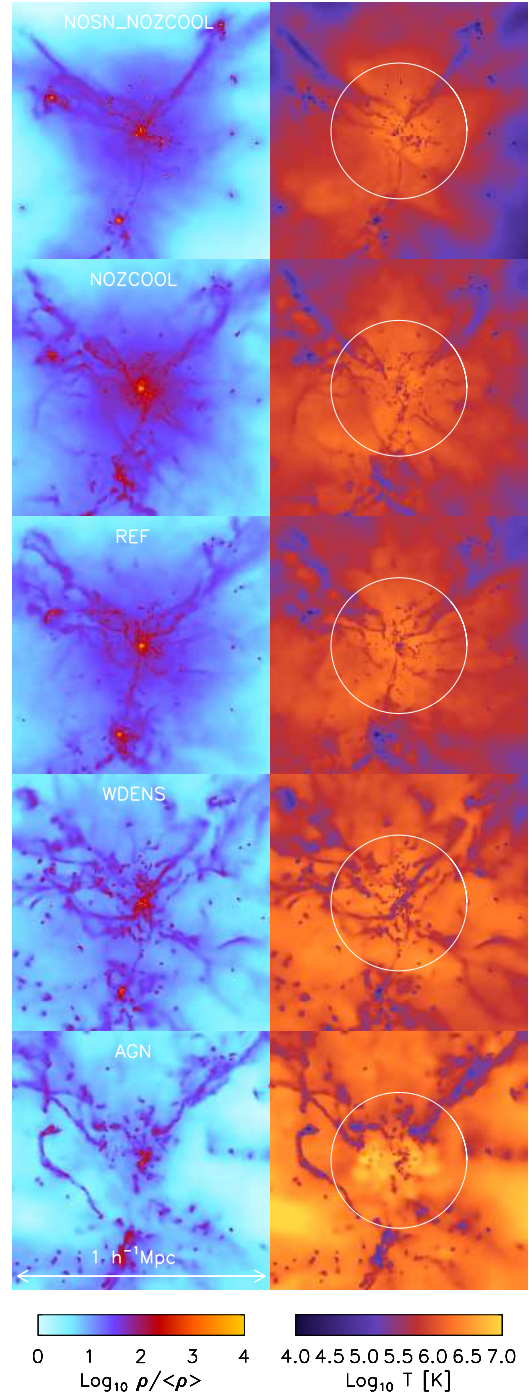}
\caption {\label{fig:diffsimzoom} Gas overdensity (left) and temperature (right) in a 1~$h^{-1}$ comoving Mpc box centred on a halo of $10^{12}$~M$_\odot$ at $z=2$. We show the same halo as was shown in the middle panel of Figure~\ref{fig:single}, but now for five different models. The white circles indicate the virial radius of the halo. From top to bottom: no SN feedback and no metal-line cooling (\emph{NOSN\_NOZCOOL}), no metal-line cooling (\emph{NOZCOOL}), reference SN feedback (\emph{REF}), density dependent SN feedback (\emph{WDENS}), reference SN feedback, and AGN feedback (\emph{AGN}). These simulations used $2\times512^3$ particles in a 25~$h^{-1}$Mpc box. The structure of the cold streams changes, but they exist in all simulations. The hot gas extends to larger radii and has a higher temperature if feedback is more efficient. The structure inside the halo is clearly different in different simulations, but cold gas is always present outside the disc.}
\end{figure}

As discussed in Section~\ref{sec:accrate}, feedback from SNe and AGN decreases the halo accretion rate, while metal-line cooling has very little effect. In this Section we will discuss the influence of cooling and feedback on the hot and cold accretion fractions at $z=2$ and $z=0$.

To investigate the influence of galactic winds driven by SN feedback, we ran simulations with no SN feedback at all and with more effective galactic wind models, that can eject gas from more massive haloes. In other simulations, cooling rates are computed assuming primordial abundances. We have also included AGN feedback in one simulation.

To illustrate the effect these different processes have on an individual halo, Figure~\ref{fig:diffsimzoom} shows the same $10^{12}$~M$_\odot$ halo as was shown in the middle panel of Figure~\ref{fig:single} for five different simulations. From top to bottom are shown: no SN feedback and no metal-line cooling (\emph{NOSN\_NOZCOOL}); no metal-line cooling (\emph{NOZCOOL}); reference SN feedback (\emph{REF}); density dependent SN feedback (\emph{WDENS}); reference SN feedback and AGN feedback (\emph{AGN}). The colour coding shows gas overdensity in the left panels and temperature in the right panels. 
The stronger the feedback (top row: weakest feedback, bottom row: strongest feedback) the more fragmented the streams become. All models predict the presence of cold, dense gas throughout much of the halo. The diffuse halo gas is heated to higher temperatures in simulations with strong SN or AGN feedback than in the reference simulation. The radius out to which gas is heated increases for strong feedback models.

\begin{figure*}
\center
\includegraphics[scale=0.62]{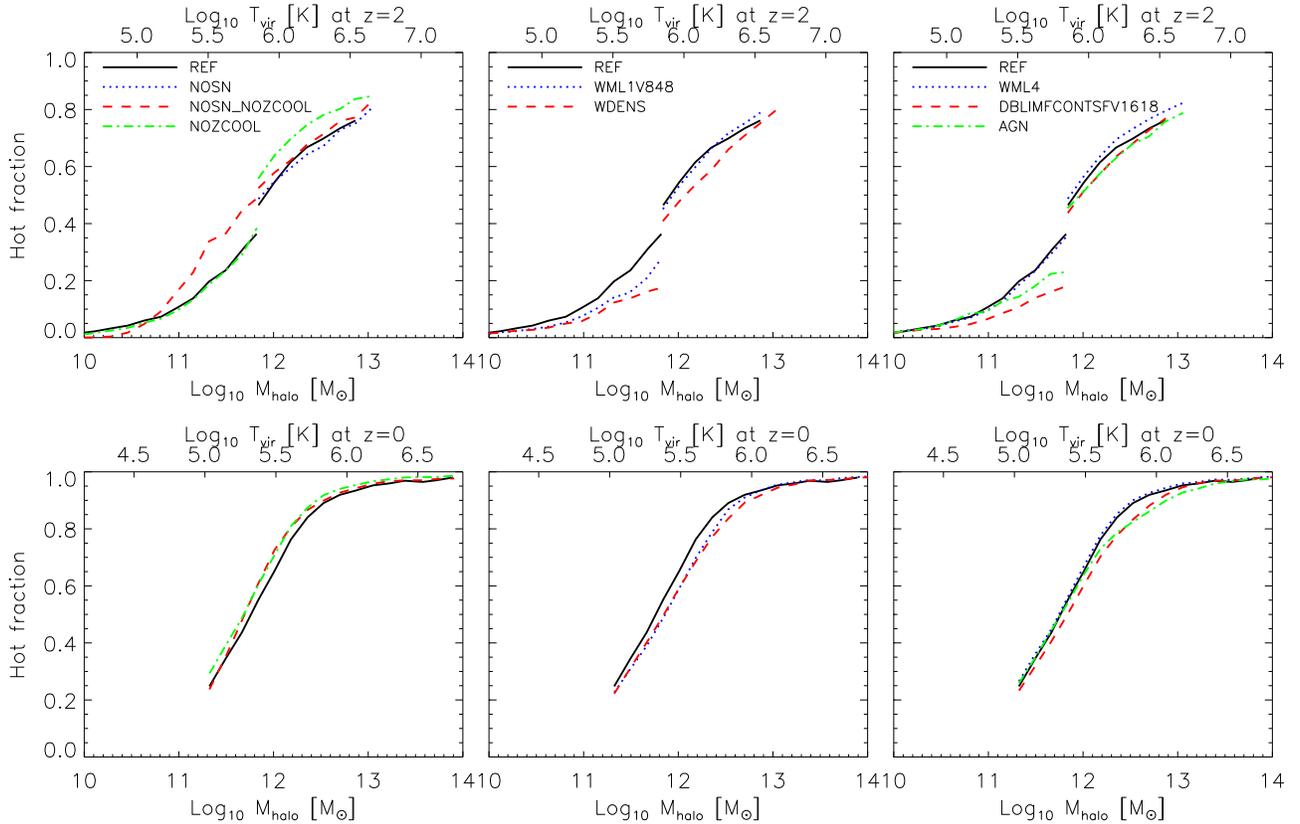}
\caption {\label{fig:diffsim} The average fraction of the gas that accreted smoothly on to haloes just before $z=2$ (top panels) and $z=0$ (bottom panels), that reached maximum past temperatures above $10^{5.5}$~K (by $z=2$ and $z=0$, respectively) is plotted as a function of total halo mass. The line styles are identical to those used in Figure~\ref{fig:diffaccrate}. The curves at the high-mass end are from simulations \emph{L100N512}. At $z=2$ the curves at the low-mass end are from simulations \emph{L025N512}, resulting in a factor 64 higher mass resolution. Each mass bin contains at least 10 haloes. At $z=2$, each halo contains at least 1000 dark matter particles. This is higher than our resolution limit for accretion on to haloes, but it removes the overlap between simulations of different resolution. The differences between the simulations are small. Efficient feedback generally reduces the hot fraction, indicating that hot gas is more vulnerable to feedback than cold gas.} 
\end{figure*}

Figure~\ref{fig:diffsim} shows how the fraction of the gas that accretes smoothly on to haloes just before $z=2$ (top panels) and $z=0$ (bottom panels) in the hot mode, i.e.\ with $T_\mathrm{max}\ge10^{5.5}$~K, depends on the physical processes that are modelled.
Even though the hot fractions are not completely converged at the low-mass end for some of the \emph{L100N512} runs, the trend with halo mass and the effect of the variations in the simulations are robust. The haloes above $M_\mathrm{halo}\approx10^{12}$~M$_\odot$ in the \emph{L025N512} runs (not plotted) show an increase in the hot fraction as steep as the \emph{L100N512} runs.
Before discussing the differences between the models, we stress that these differences are small. The fraction of the gas that accretes on to a halo of a given mass and at a given redshift in the hot mode can thus be robustly estimated. It is insensitive to uncertainties in the baryonic physics, such as radiative cooling and feedback from star formation and AGN.

We first focus on the left panels, which compare simulations with and without metal-line cooling and with and without SN feedback. For high-mass haloes the fraction of gas accreting in the hot mode is a little bit higher without metal-line cooling, because the gas will reach higher temperatures if the cooling rates are lower. The effect is, however, small. 

The hot fraction depends only slightly on the specific feedback model used if the energy per unit stellar mass is kept fixed, as can be seen from the middle panels of Figure~\ref{fig:diffsim}. The hot fraction is a bit lower for the model with the most effective feedback (\emph{WDENS}).

The right panels show that AGN feedback decreases the hot fraction, though the effect is not large and limited to high-mass ($\gg 10^{11}$~M$_\odot$) haloes. More effective stellar feedback as a result of a top-heavy IMF in starbursts (\emph{DBLIMFCONTSFV1618}) reduces the hot fraction more for low-mass $\lesssim10^{12}$~M$_\odot$ haloes.
These results suggest that the hot accretion mode is slightly more affected by feedback than the cold mode. Because the hot gas is less dense than the cold gas and because it spans a greater fraction of the sky as seen by the galaxy, it is more likely to be affected by feedback. It has been argued by several authors that AGN feedback would therefore couple mostly to the hot gas \citep{Keres2005, Dekel2006}. Our results show that this effect is small, although we will show below that it is important for accretion on to galaxies residing in high-mass haloes at low redshift. The main conclusion is that the fraction of the gas that accretes on to haloes in the hot mode is insensitive to feedback from SNe and AGN.

\section{Hot and cold accretion on to galaxies} \label{sec:sf}

A significant fraction of the gas that accretes on to haloes may remain in the hot, low-density halo. This gas never cools and it will not get into the inner galaxy and contribute to the SFR. This diffuse gas may be easily pushed out of the halo by galactic winds. It is therefore of interest to look not only at the gas that accretes on to haloes, but also at the gas that actually accretes on to galaxies. After all, it is only the gas that is accreted on to galaxies that is available for star formation.

Gas is by definition cold when it accretes on to a galaxy, because it must cool below $T=10^5$~K to be able to join the ISM.
However, by using $T_\mathrm{max}$ we are probing the entire temperature history of the gas and not just the temperature at the time of accretion. Gas accreting on to a galaxy in the hot mode has been hot in the past (usually when it accreted on to the halo), but was able to cool down and reach the central galaxy. Gas accreting on to a galaxy in the cold mode has never been hot in the past. 

\begin{figure}
\center
\includegraphics[scale=0.5]{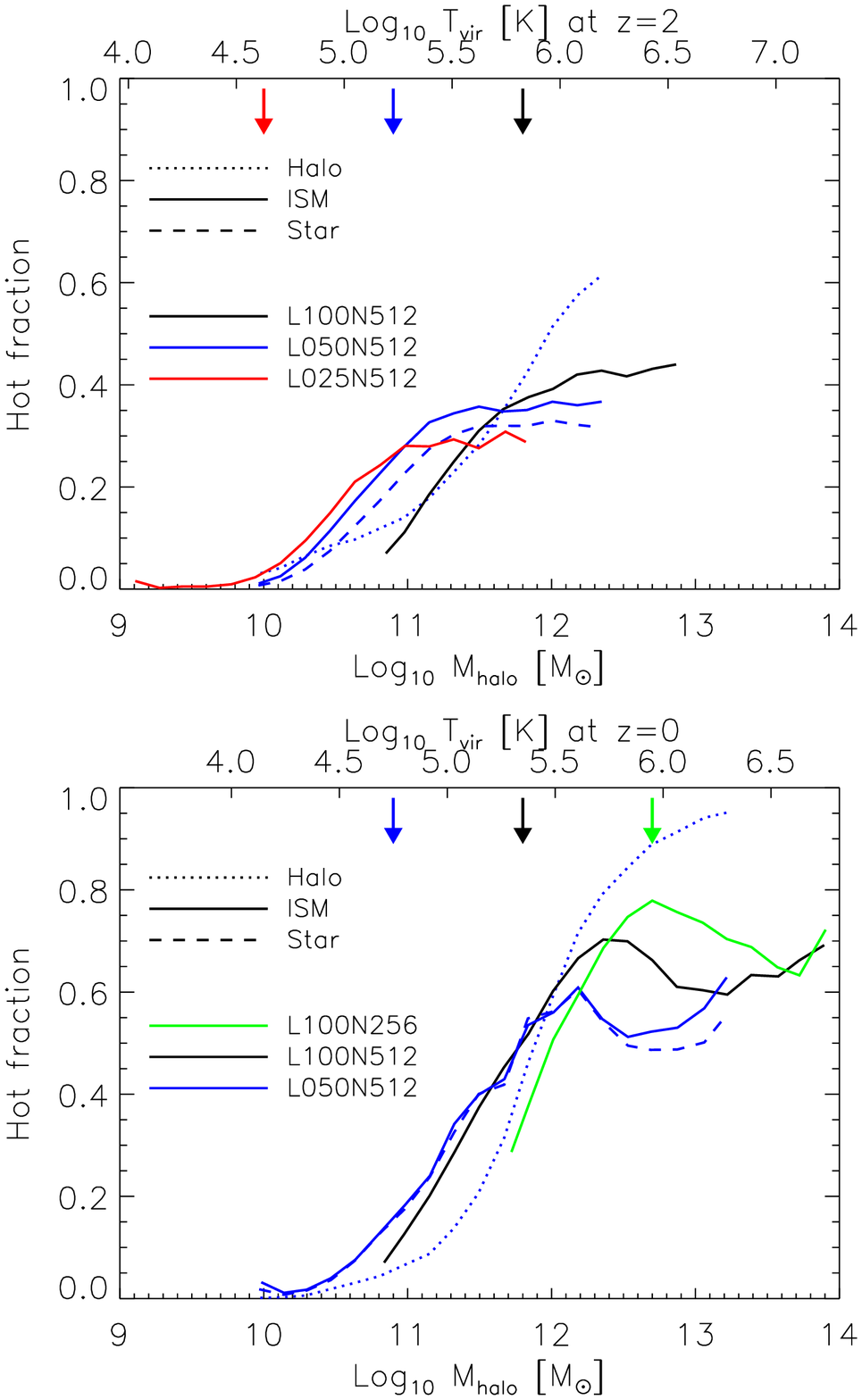}
\caption {\label{fig:haloeosstar} The dotted curve shows the average fraction of gas smoothly accreted on to haloes at $z=2$ (top panel) and $z=0$ (bottom panel), that has maximum past temperatures above $T_\mathrm{max}=10^{5.5}$~K. The solid curves show the average fraction of the gas smoothly accreted on to the ISM with maximum past temperatures above $T_\mathrm{max}=10^{5.5}$~K. The dashed curves show the fraction of the stars formed in the same redshift intervals from gas with maximum past temperatures above $T_\mathrm{max}=10^{5.5}$~K. These reference simulations span a factor of 64 in mass resolution in each panel. Each mass bin contains at least 10 haloes. Arrows correspond to the adopted resolution limit for accretion on to galaxies. For high-mass haloes, the hot fractions for accretion on to haloes and galaxies diverge. The hot fraction for recently formed stars follows the same trend as for accretion on to the ISM.}
\end{figure}
In Figure~\ref{fig:haloeosstar} we show the hot fraction for accretion on to haloes (dotted curve), accretion on to the ISM (solid curve), and for stars formed (dashed curve) just before $z=2$ (top panel) and just before $z=0$ (bottom panel) for the 50~$h^{-1}$Mpc reference simulation. 
To illustrate the convergence, we also show the hot fraction for accretion on to the ISM for simulations with different resolutions.
We show all results down to halo masses corresponding to 100 dark matter particles, i.e.\ 1~dex below our resolution limit.

For high-mass haloes ($M_\mathrm{halo}\gtrsim10^{12}$~M$_\odot$) hot mode accretion is less important for accretion on to the ISM, and therefore for star formation, than for accretion on to haloes.
For $T_\mathrm{vir}\gtrsim10^{6}$~K the hot fraction remains approximately constant with mass, even though the hot fraction for halo accretion becomes larger. The lower hot fractions arise because the temperature of the hot gas increases with the virial temperature and for these temperatures hotter gas has a longer cooling time \citep[e.g.][]{Wiersma2009a}, making it less likely to enter the galaxy. In reality, cold, dense clouds may be disrupted more easily than in SPH simulations, which could push the hot fraction up \citep{Agertz2007}.

On the other hand, for low-mass haloes ($T_\mathrm{vir}\lesssim10^{5.5}$~K) the hot fraction for accretion on to galaxies is higher than for accretion on to haloes. The virial temperature of these haloes is so low that the maximum temperature will only be above $10^{5.5}$~K if the gas was heated by SN feedback. This can happen after the gas has accreted on to the halo, but before the gas joins the ISM, explaining the higher hot fraction for accretion on to galaxies. Indeed, \citet{Oppenheimer2010} have shown that the re-accretion of gas that has been ejected by galactic winds can be important. The hot fraction for low-mass haloes is lower in simulations without feedback, confirming our interpretation (see Figure~\ref{fig:diffsimeos}).

As expected, for recently formed stars the hot fraction is comparable to that for gas that recently accreted on to the ISM, although it is a bit lower. It is slightly lower because of the time delay between accretion and star formation. The gas from which the stars were formed was typically accreted at higher redshift and on to lower-mass haloes, which corresponds to lower hot fractions.

Our lowest resolution simulations underestimate the hot fraction somewhat at their low-mass ends, but this regime is excluded by our adopted resolution limit, as indicated by the arrows. However, for $z=0$ the results are also not fully converged at the high-mass end ($M_\mathrm{halo}>10^{12}$~M$_\odot$). For high-mass haloes the hot fraction decreases with increasing resolution, as we also saw for accretion on to haloes (see Figure~\ref{fig:ratio2p0}), although the effect is much larger for accretion on to galaxies. This is likely because higher densities can be reached with higher resolution. Cold mode accretion may thus be somewhat more important for fuelling massive galaxies than our simulations suggest.

\begin{figure}
\center
\includegraphics[scale=0.5]{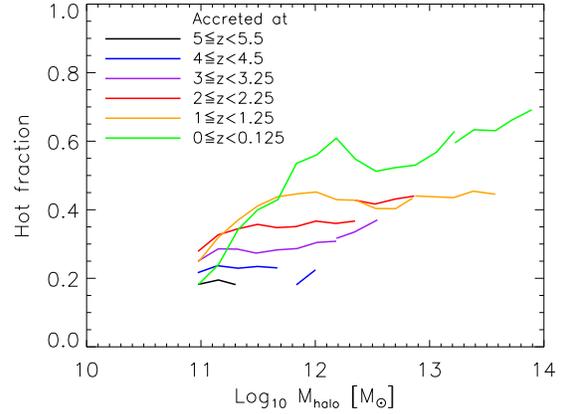}
\caption {\label{fig:eosz} Hot fraction against halo mass for accretion on to the ISM just before $z=5$, 4, 3, 2, 1, and 0 shown from bottom to top by the curves in black, blue, purple, orange, red, and green, respectively. The curves at low halo masses are obtained from model \emph{REF\_L050N512}. At the high-mass end we have added curves for \emph{REF\_L100N512} to extend the dynamic range. Each mass bin contains at least 10 haloes. The hot fraction increases with redshift, except for haloes with virial temperatures which fall below $10^{5.5}$~K, the value we use to separate hot and cold accretion ($M_\mathrm{halo}\lesssim10^{11.5}$~M$_\odot$ and $z\lesssim1$). 
The hot fraction varies less strongly with halo mass than for accretion on to haloes.}
\end{figure}
The evolution of the hot fraction of gas accreted on to the ISM is shown in Figure~\ref{fig:eosz} from $z=5$ (bottom, black curve) to $z=0$ (top, green curve) as a function of halo mass. 
The trend with redshift is the same as for accretion on to haloes (see Figure~\ref{fig:z5to0}). The hot fraction increases with decreasing redshift for all halo masses, except when the virial temperatures fall below $10^{5.5}$~K, the value we use to separate the hot and cold accretion modes, which happens for $M_\mathrm{halo}\lesssim10^{11.5}$~M$_\odot$ at $z\lesssim1$. The hot fraction for accretion on to galaxies increases less steeply with halo mass than for accretion on to haloes. This is due to the fact that the cooling time of the hot gas increases for higher-mass haloes, making it less likely that hot gas reaches the central galaxy.

For haloes with $T_\mathrm{vir}\gtrsim10^6$~K, the hot fraction of gas that reaches the ISM is much lower than the hot fraction of all the gas that accretes on to the halo. At high redshift ($z\gtrsim4$) cold accretion is the dominant mode for feeding galaxies. At lower redshift ($z\lesssim2$) hot and cold accretion are comparable. Except for high-mass haloes at low redshift ($M_\mathrm{halo}>10^{13}$~M$_\odot$ at $z=0$), hot mode accretion is less important for feeding the central galaxy than cold mode accretion. It is always less important for the growth of galaxies than it is for the growth of haloes, though it is never negligible. We found the hot fraction of recently formed stars to be slightly lower, because it takes some time to convert the gas into stars. Cold mode accretion is most important for the total build-up of stellar mass in galaxies.

We have not included galaxy mergers. Including mergers preferentially brings in gas accreted in the cold mode, because that gas was already part of the ISM. This reduces the hot fraction slightly for high-mass haloes at low redshift. Cold mode accretion is therefore also the main mode for galaxy growth in this case.

\subsection{Effects of physical processes} 

We showed the effect of feedback, metal-line cooling, and cosmology on galaxy accretion rates in Section \ref{sec:accrate}. Feedback reduces these rates, while including metal-line cooling increases them. In this Section we discuss their consequences for the relative importance of hot and cold accretion.

\begin{figure*}
\center
\includegraphics[scale=0.62]{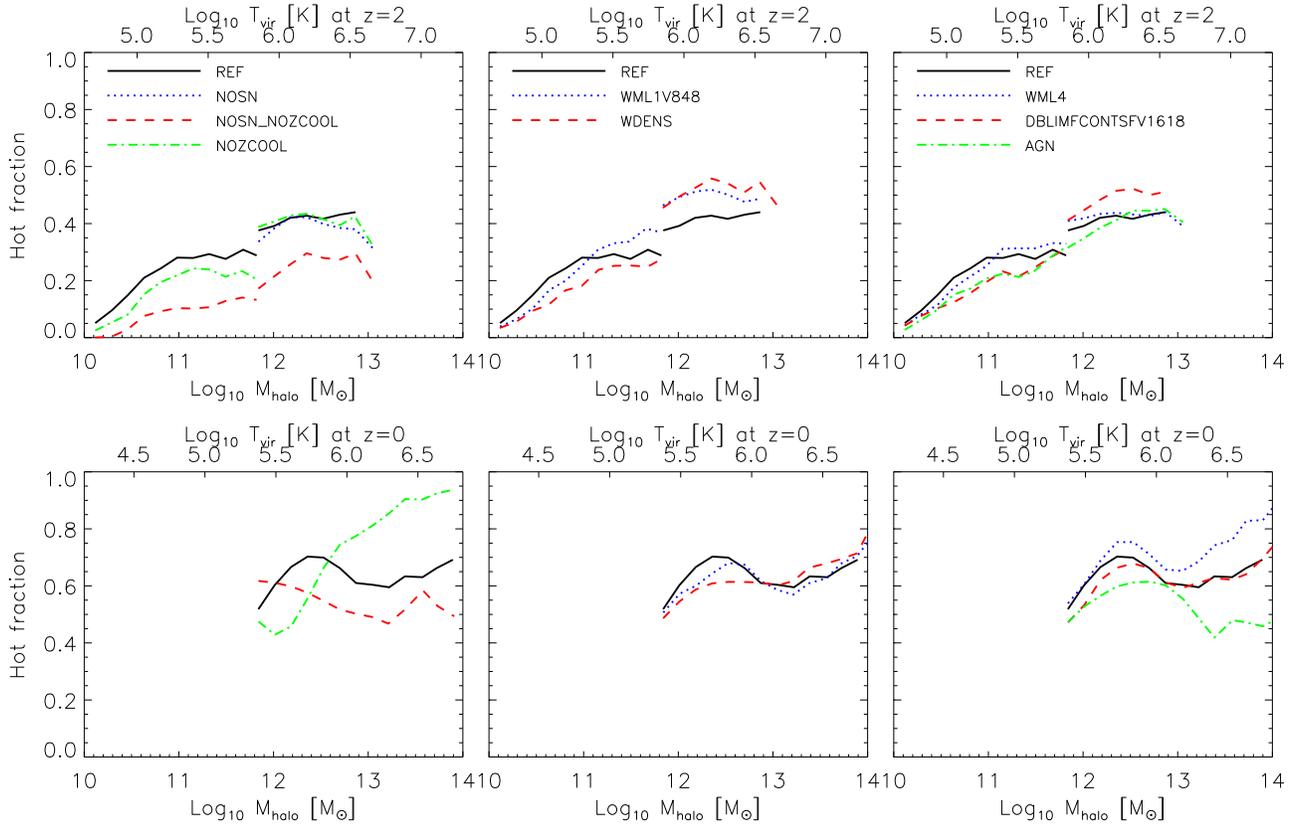}
\caption {\label{fig:diffsimeos} The average fraction of the gas accreting on to the ISM at $z=2$ (top panels) and $z=0$ (bottom panels), that has maximum past temperatures $T_\mathrm{max}\ge10^{5.5}$~K. The line styles are identical to those used in Figure~\ref{fig:diffsim}. The curves at the low and high-mass ends are from simulations \emph{L025N512} and \emph{L100N512}, respectively, which differ by a factor 64 in mass resolution. Each mass bin contains at least 10 haloes. Differences in results from different simulations are mostly small, except for $M_\mathrm{halo}> 10^{13}$~M$_\odot$ at $z=0$ if metal-line cooling is ignored or AGN feedback is added.}
\end{figure*}
Figure~\ref{fig:diffsimeos} shows the hot fraction of the gas that reaches the ISM between $z=2.25$ and $z=2$ as a function of halo mass for the same simulations as were shown for accretion on to haloes in Figure~\ref{fig:diffsim}. Even though the hot fractions are not completely converged for the \emph{L100N512} runs, the trend with halo mass and the effect of the variations in the physics are robust. As was the case for accretion on to haloes, the trends are the same for all simulations, although the differences between models are larger for accretion on to the ISM than for accretion on to haloes. At $z=2$, the hot mode is less important for accretion on to the ISM, and therefore less important for star formation, than the cold mode. For the halo mass range shown at $z=0$ ($\gtrsim10^{12}$~M$_\odot$), the hot mode is slightly more important than the cold mode.

We can observe the difference between simulations with and without metal-line cooling and with and without SN feedback in the left panels. The simulation without metal-line cooling (\emph{NOZCOOL}) has a slightly smaller hot fraction than the simulation with metal-line cooling (\emph{REF}) at $z=2$ and for $M_\mathrm{halo}\lesssim10^{12.5}$~M$_\odot$ at $z=0$. Because the cooling times are longer in the absence of metal-line cooling, less hot gas is able to reach the high densities that define the ISM. For accretion on to haloes we found the opposite effect (see Figure~\ref{fig:diffsim}), because the lower cooling rate increases the maximum temperature reached by gas accreted on to haloes and hence increases the corresponding hot fraction. For high mass haloes ($M_\mathrm{halo}\gtrsim10^{13}$~M$_\odot$) at $z=0$, the hot fraction is much higher in the absence of metal-line cooling. In Figure~\ref{fig:diffaccrateeos} we have seen that metal-line cooling strongly increases the accretion rates on to galaxies in lower mass galaxies. Thus, without metal-line cooling, less hot gas accretes on to low-mass haloes, leaving more hot gas to accrete on to their higher mass descendants.

We can compare the simulation without metal-line cooling (\emph{NOZCOOL}) to the one without SN feedback and without metal-line cooling (\emph{NOSN\_NOZCOOL}). In the simulations with feedback a higher fraction of the gas accreted on to the ISM has been hot, presumably because some of the gas accreted cold on to haloes was heated by outflows before it joined the ISM (for the last time). At the high-mass end at $z=0$ the differences between the curves parallels those in the specific accretion rates (Figure~\ref{fig:diffaccrateeos}). This suggests that here SN feedback increases the hot fraction because it prevents the accretion of hot gas on to lower mass progenitors, leaving more gas available to cool in high-mass galaxies, where the SN feedback is inefficient. 
The effect of SN feedback on the hot fraction at $z=2$ is much smaller if metal-line cooling is included (\emph{NOSN}). This may indicate that the increase in the cooling rates due to the metals carried by the winds compensates for the extra shock heating.

The middle panels show the result for simulations with different values for the SN feedback parameters, but the same amount of SN energy per unit stellar mass formed. The hot fractions are similar, though slightly higher if the feedback model is more efficient (\emph{WML1V848} for $M_\mathrm{halo}>10^{11}$~M$_\odot$ and \emph{WDENS} for $M_\mathrm{halo}>10^{12}$~M$_\odot$ at $z=2$).

Results for simulations with a prescription for SN feedback that use more energy and for a model including AGN feedback are shown in the right panels. Perhaps surprisingly, at $z=2$ the simulation including AGN feedback (\emph{AGN}) gives results similar to the reference simulation. The accretion rate on to the galaxy is suppressed by up to 1~dex, as can be seen in Figure~\ref{fig:diffaccrateeos}, but the hot fraction is almost the same. At $z=0$ the hot fraction is somewhat lower, though still above 40\%.

In Figure~\ref{fig:diffsim} we showed that the hot fraction for the gas accreting on to haloes does not change much if we vary the prescriptions for feedback and radiative cooling. To first order, the same conclusion is true for the gas that accretes on to the ISM and thus becomes star forming, although the differences are larger. At $z=2$ the hot fraction decreases significantly if both SN feedback and metal-line cooling are turned off. For high-mass galaxies ($M_\mathrm{halo}> 10^{13}$~M$_\odot$) at $z=0$ the importance of the hot mode increases substantially if metal-line cooling is excluded and decreases significantly if AGN feedback is added. The specific implementation of the SN feedback does not have a large effect on the fraction of the gas accreting on to galaxies that has a maximum past temperature $T_\mathrm{max}\ge10^{5.5}$~K.

\section{Comparison with previous work} \label{sec:discuss}

We can compare our results to the pioneering work of \citet{Keres2005,Keres2009a}, and \citet{Ocvirk2008}. All these studies used 250,000~K as the critical temperature to separate hot and cold accretion. We used $10^{5.5}=316,228$~K, but our results would have been very similar if we had used 250,000~K. 

\citet{Ocvirk2008} use an adaptive mesh refinement (AMR) code and include metal enrichment and cooling, but only weak SN feedback and no AGN feedback. They use only a single simulation, whose resolution is similar to our \emph{L025N512} runs. Their simulation was only run to $z\approx1.5$, so we can only compare to our high-redshift results. Their box size of 50~$h^{-1}$Mpc is too small to sample haloes with mass $\gtrsim10^{13}$~M$_\odot$. As their simulation is not Lagrangian, they cannot trace the gas back in time, and were forced to separate the cold and hot modes based on current temperatures. 

For $z=2-3$ they predict that the hot fraction for accretion on to haloes reaches 0.5 at $M_\mathrm{halo}\sim 10^{11.5}$~M$_\odot$, whereas we find $M_\mathrm{halo}\sim10^{12}$~M$_\odot$. For higher redshifts we predict somewhat higher hot fractions. 
We predict a gradual increase in the transition mass with redshift, but see no sign of the sudden change between $z=3$ and 4 that they found. 

For accretion on to galaxies, which they measure at 0.2 virial radii, \citet{Ocvirk2008} also do not predict a sudden change with redshift. As the gas at this radius is not necessarily star forming and has not necessarily cooled down to $T\le10^5$~K, it is not possible to make a completely fair comparison. They find no change with redshift, whereas we find strong evolution. The hot fraction increases up to their highest halo mass, reaching values close to unity for $10^{12.5}$~M$_\odot$. In our simulations, however, the hot fraction for accretion on to galaxies is approximately constant and never exceeds 0.4 at $z\ge2$. As we are using the SPH technique, it is possible that our simulations suffer from in-shock cooling, which would lead to an underestimate of the hot fraction. However, we find that the hot fraction in fact decreases with increasing resolution, the opposite of what we would expect if this were an important effect. Perhaps the difference is due to the fact that we do not update the maximum past temperature once the gas has become star forming. If the accretion shock on to the galaxy happens after that time, we would underestimate the maximum temperature. This is, however, consistent with our aims, as we are interested in the maximum temperature reached \emph{before} the gas entered the galaxy. 

\citet{Keres2005,Keres2009a} use SPH codes. They ignore metal-line cooling and do not include feedback from SNe or AGN. The resolution of their main simulation is comparable to our \emph{L100N512} runs, but their simulation volume is more than eight times smaller. Their mass resolution is nearly two orders of magnitudes worse than for our \emph{L025N512} runs. They identify galaxies using a different group finder, \textsc{skid}, which links bound stars and dense, cold gas ($\rho/\langle\rho_\mathrm{baryon}\rangle>1000$ and $T<30,000$~K). These groups are considered to be galaxies. In contrast to our work, they include accretion on to both central galaxies and satellites, which could also lead to somewhat different results. Accretion on to haloes was not investigated. 

Contrary to our results and those of \citet{Keres2009a}, \citet{Keres2005} find that the hot fraction continues to increase with halo mass and that there is no significant evolution. However, using the same method, \citet{Keres2009a} find much lower hot fractions than \citet{Keres2005}. They find that this difference is mostly due to the fact that they switched to the entropy conserving formulation of SPH of \citet{Springel2002}, which prevents overcooling due to artificial phase mixing and which we have used as well. We will therefore only compare with \citet{Keres2009a} (and only for accretion on to galaxies). 

At high redshift ($z=4$) we find similar results for the hot fractions as \citet{Keres2009a}, although ours are slightly higher, which is due to the fact that we include SN feedback. At $z\le2$ they find that the hot fraction first increases, reaches a maximum around $M_\mathrm{halo}=10^{12}$~M$_\odot$, after which it decreases. We find that the hot fraction varies less strongly with halo mass and that it remains approximately constant at the high mass end.

Our results are also in qualitative agreement with \citet{Brooks2009}, who used SPH simulations (without metal-line cooling and AGN feedback) of a few individual galaxies with $M_\mathrm{halo}\lesssim10^{12}$~M$_\odot$. A detailed comparison is difficult as their sample is too small to obtain statistics and because they used a more complicated criterion to separate the different accretion modes.

\section{Conclusions} \label{sec:conclude}

Before summarizing and discussing our findings, we list our main conclusions:
\begin{itemize}
\item To first order, the rate of gas accretion on to haloes follows that of dark matter. Except for low-mass haloes ($M_\mathrm{halo}\ll10^{11}$~M$_\odot$), feedback changes these rates only slightly.
\item Except for groups and clusters, gas accretes mostly smoothly (i.e.\ not through mergers with mass ratios greater than 1:10).
\item The rate at which gas accretes on to galaxies is set by radiative cooling, which is sensitive to the abundance of heavy elements, and by feedback from SNe and AGN. Galactic winds driven by star formation increase the halo mass at which the central galaxies grow the fastest by about two orders of magnitude to $M_\mathrm{halo}\sim10^{12}$~M$_\odot$. 
\item The signs of the effects of feedback and metal-line cooling on gas accretion can change with halo mass.
\item Gas accretion is bimodal, with maximum past temperatures either of order the virial temperature or $\lesssim10^5$~K. Both modes can be present in a single halo, the cold mode being most prominent in filaments.
\item The fraction of gas accreted in the hot mode (i.e.\ maximum past temperature $T_\mathrm{max}\ge10^{5.5}$~K), increases with halo mass and with decreasing redshift.
\item For accretion on to haloes, the relative importance of the hot and cold modes is is robust to changes in the feedback prescriptions. Cold and hot accretion dominate for $M_\mathrm{halo}\ll 10^{12}$~M$_\odot$ and $M_\mathrm{halo}\gg10^{12}$~M$_\odot$, respectively.
\item For accretion on to galaxies the cold mode is always significant and the relative importance of the two accretion modes is much more sensitive to feedback and cooling than is the case for halo accretion. 
\item On average, most of the stars present in any mass halo at any redshift were formed from gas accreted in the cold mode, although the hot mode contributes typically over 10\% for $M_\mathrm{halo}\gtrsim10^{11}$~M$_\odot$ (see Figure~\ref{fig:cumstar} below). 
\end{itemize}

While the rate at which dark matter accretes on to haloes can be reliably calculated, the situation is rather more complicated for gas. Gas may be heated through accretion shocks, but can also radiate away its thermal energy. This cooling rate is, however, strongly affected by contamination with metals blown out of galaxies. Such galactic winds driven by star formation or accreting supermassive black holes may also directly halt or reverse the accretion, which may in turn cause gas elements to be recycled multiple times. 

Pressing questions include: What fraction of the gas accreting on to haloes experiences a shock near the virial radius?
How does this fraction vary with halo mass and redshift? What fraction of the gas that falls into a dark matter halo accretes on to a galaxy and how does this vary with mass and redshift? How do processes like metal-line cooling and feedback from star formation and AGN affect gas accretion?
We addressed these questions by analysing a large number of cosmological, hydrodynamical simulations from the OWLS project \citep{Schaye2010}. By repeating the simulations many times with
varying parameters, we investigated what physical processes drive the
accretion of gas on to galaxies and haloes. For each physical model we combined at least two $2\times512^3$ particle simulations in order to cover a dynamic range of about 4 orders of magnitude in halo mass ($M_\mathrm{halo}\sim10^{10}-10^{14}$~M$_\odot$). \citet{Schaye2010} have shown that the simulation with AGN feedback is able to reproduce the steep slope of the observed star formation rate density at $z<2$. This is a significant improvement over previous simulations. In a future paper we will discuss the contributions of hot and cold accretion to the cosmic star formation history.

Except for $M_\mathrm{halo}\gg10^{13}$~M$_\odot$ at low redshift, mergers with mass ratios exceeding 1:10 contribute $\lesssim10$\% of the total accretion on to haloes. The growth of haloes is thus dominated by smooth accretion. 
The specific rate of smooth gas accretion on to haloes is close to that for dark matter accretion, particularly at higher redshifts. It decreases with time and increases with halo mass. The increase with halo mass is, however, gradual and the gradient decreases with increasing halo mass. To first order, the halo accretion rate scales linearly with the halo mass. 

The rate of accretion on to haloes is relatively insensitive to the inclusion of metal-line cooling. Efficient feedback can reduce the halo accretion rates by factors of a few. In particular, for $z=2$ we find that SN feedback reduces the halo accretion rate of low-mass haloes ($M_\mathrm{halo}\sim 10^{10}$~M$_\odot$) by a factor of about four, but is typically not efficient for $M_\mathrm{halo}\gtrsim10^{13}$~M$_\odot$.

As is the case for accretion on to haloes, accretion on to galaxies is mostly smooth. Clumpy accretion, which in this case we defined as accretion of material that already had densities $n_\mathrm{H}\ge0.1$~cm$^{-3}$ at the previous snapshot, only becomes of comparable importance as smooth accretion for $M_\mathrm{halo}\gtrsim10^{13}$~M$_\odot$. 

While the specific accretion rate on to haloes increases slowly with halo mass over the full range spanned by our simulations, the specific accretion rate on to galaxies increases rapidly for $M_\mathrm{halo}\ll10^{12}$~M$_\odot$ and drops quickly for $M_\mathrm{halo}\gg10^{12}$~M$_\odot$. The halo mass at which the specific accretion rate on to galaxies peaks is sensitive to feedback from SNe and AGN. Without SN feedback, the specific accretion rate peaks at
$M_\mathrm{halo}\sim10^{10}$~M$_\odot$, where it exceeds the rate predicted by runs that do include SN feedback by an order of magnitude. For higher mass haloes ($M_\mathrm{halo}\gg10^{11}$~M$_\odot$), on the other hand, SN feedback tends to increase the specific accretion rates by about a factor of two, either because of the increased importance of recycling (i.e.\ the same gas can be re-accreted after it has been ejected, see \citealt{Oppenheimer2010}) or because without SN feedback the accreted gas would already have been consumed in lower mass progenitor galaxies. AGN feedback can strongly reduce the accretion rates on to galaxies. This sensitivity to feedback is in contrast to accretion on to haloes. In the absence of metal-line cooling, the peak in the specific accretion rate is
less pronounced. Hence, cooling and feedback set the efficiency of galaxy formation by controlling what fraction of the gas that accretes on to haloes is able to accrete on to galaxies. 

The rate of accretion on to galaxies is smaller than the accretion rate on to haloes. The difference between the two rates increases with time and is minimal for $M_\mathrm{halo}\sim10^{12}$~M$_\odot$. For this mass the difference increases from a factor of two at $z=2$ to a factor of four at $z=0$.

Tracing back in time, most gas particles that reside in haloes at $z=0$ reached their maximum temperature around, or shortly after, the time they were first accreted on to a halo. In the presence of a photo-ionising background, essentially all gas accreting smoothly on to haloes has been heated to $\gtrsim10^5$~K. For haloes with virial temperatures $\gg10^5$~K the probability distribution for the maximum past temperature reached by gas accreted on to haloes is bimodal. The low temperature peak at $\sim10^5$~K represents gas that accreted predominantly through large-scale filaments (but still smoothly). The higher densities in the filaments enable efficient cooling which in turn prevents the establishment of a stable accretion shock \cite[e.g.][]{Keres2005, Dekel2006}. The high-temperature peak coincides with the virial temperature of the halo and is reached through an accretion shock near the virial radius.  

We separated these two modes of smooth gas accretion according to the maximum past temperature of the gas, which we updated for each particle at each time step, and referred to them as cold ($T_\mathrm{max}<10^{5.5}$~K) and hot ($T_\mathrm{max}\ge10^{5.5}$~K) accretion, respectively. For high-mass haloes (virial temperatures $\gtrsim10^6$~K), the hot fraction, i.e.\ the fraction of gas that is accreted in the hot mode, reflects the fraction of the accreted gas that has experienced an accretion shock near the virial radius. We
emphasized, however, that for gas accreted on to haloes with virial
temperatures $\lesssim10^{5.5}$~K ($M_\mathrm{halo}\lesssim10^{11.5}$~M$_\odot$ at $z=2$ and $M_\mathrm{halo}\lesssim10^{12}$~M$_\odot$ at $z=0$) gas accreted in the cold mode (according to our definition) may also have gone through a virial shock. 

The fraction of the gas that is accreted on to haloes and galaxies of a given mass in the cold mode typically increases with redshift. This is because radiative cooling is more efficient at high redshift, where the gas densities are higher (and the cooling time is more sensitive to the density than the dynamical and Hubble times). 

For accretion on to haloes, the relative importance of the two modes is mostly determined by the halo mass. At $z=0$ the hot fraction increases from less than 0.1 at $M_\mathrm{halo}\sim10^{11}$~M$_\odot$ to more than 0.9 at $M_\mathrm{halo}\sim10^{13}$~M$_\odot$. For all redshifts for which we have sufficient statistics ($z<5$), the hot fraction reaches 0.5 for $M_\mathrm{halo}\sim10^{12}$~M$_\odot$, in reasonable agreement with the predictions of \citet{Birnboim2003} and \citet{Dekel2006} for metal enriched gas at the virial radius and with previous simulations \citep{Keres2005, Ocvirk2008, Keres2009a}.
Contrary to accretion on to haloes, the hot fraction for gas accreted on to galaxies is only weakly dependent on halo mass. In particular, the hot fraction for accretion on to galaxies is nearly constant for $M_\mathrm{halo}\gtrsim10^{12}$~M$_\odot$ and remains below 0.5, except at $z\approx0$. 

For accretion on to haloes, the relative importance of the cold and hot modes is mostly insensitive to the inclusion of metal-line cooling or feedback from SNe and AGN. For accretion on to galaxies the effects are stronger, though generally still weak. The main exceptions are galaxies in high-mass haloes $M_\mathrm{halo}>10^{13}$~M$_\odot$ at low redshift. For these systems the hot mode becomes much more important in the absence of metal-line cooling (because less gas has cooled on to their progenitors) and substantially less important if AGN feedback is considered (because it efficiently removes hot gas). 

As expected, the hot fraction for the gas that is converted into stars is very similar to that of the gas that accretes on to galaxies. As illustrated in Figure~\ref{fig:cumstar}, where we show the fraction of the total stellar mass formed from gas which has reached $T_\mathrm{max}\ge10^{5.5}$~K in its past. For the full range of redshifts and halo masses probed by our simulations, the majority of stars are formed from gas accreted in the cold mode. 

Both the comparison of our different models and the comparison with \citet{Ocvirk2008} suggest that the conclusions regarding accretion on to haloes are robust, at least for $z<4$. The gas accretion rate is similar to that of the dark matter and dominated by smooth accretion. The hot mode changes from being negligible at $M_\mathrm{halo}\lesssim10^{11}$~M$_\odot$ to strongly dominant at $M_\mathrm{halo}\gtrsim10^{13}$~M$_\odot$. Contrary to accretion on to haloes, the mode and particularly the rate of gas accretion on to galaxies, and hence the provision of fuel for star formation, is affected by uncertain processes such as metal-line cooling and especially feedback from star formation and AGN. This is reflected in the large differences in the predictions by different groups. It seems clear, however, that the hot mode is much less important for the growth of galaxies than it is for the growth of haloes.

We conclude that the rate and manner in which gas accretes on to haloes is an important ingredient of models for the formation and evolution of galaxies and that this process can be understood using simple physics. However, halo accretion gives by itself little insight into the rate and mode through which gas accretes on to galaxies. 
To understand galaxy formation, it is crucial to consider feedback processes, such as metal enrichment and outflows driven by SNe and AGN. 

\begin{figure}
\center
\includegraphics[scale=0.5]{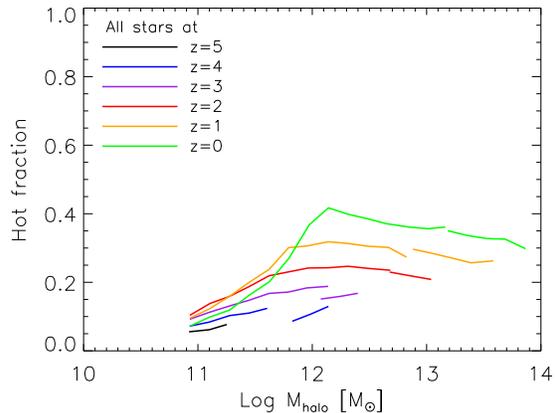}
\caption {\label{fig:cumstar} Fraction of the stellar mass at the indicated redshifts, that formed from gas that had earlier reached temperatures exceeding $T_\mathrm{max}\ge10^{5.5}$~K, as a function of the host halo mass at the corresponding redshifts for the simulation \emph{REF\_L050N512}. We added the highest mass bins from \emph{REF\_L100N512} to extend the dynamic range. Each mass bin contains at least 10 haloes. At any time and for any halo mass, hot mode accretion contributed less than 40\% to the total stellar mass.}
\end{figure}

\section*{Acknowledgements}

We would like to thank Ben Oppenheimer and Volker Springel for comments on an earlier version of the manuscript and all the members of the OWLS team for valuable discussions. We would also like
to thank the anonymous referee for constructive comments. The simulations presented here were run on Stella, the LOFAR BlueGene/L system in Groningen, on the Cosmology Machine at the Institute for Computational Cosmology in Durham as part of the Virgo Consortium research programme, and on Darwin in Cambridge. This work was sponsored by the National Computing Facilities Foundation (NCF) for the use of supercomputer facilities, with financial support from the Netherlands Organization for Scientific Research (NWO), also through a VIDI grant, and from the Marie Curie Initial Training Network CosmoComp (PITN-GA-2009-238356).

\bibliographystyle{mn2e}
\bibliography{accretion}

\bsp

\label{lastpage}

\end{document}